%% file: hertztoda.tex
\documentclass[11pt]{article}
\usepackage{amsmath,amssymb,graphicx,mathpazo,times,fullpage}
\usepackage[colorlinks=true,citecolor=blue]{hyperref}

\makeatletter
\renewcommand\section{\@startsection {section}{1}{\z@}%
  {-2ex \@plus -1ex \@minus -.2ex}%
  {1ex \@plus.1ex}%
  {\normalfont\bf\sffamily}}
\renewcommand\subsection{\@startsection{subsection}{2}{\z@}%
  {-1.75ex\@plus -0.4ex \@minus -.2ex}%
  {0.6ex \@plus .1ex}%
  {\normalfont\small\bf\sffamily}}
\renewcommand\subsubsection{\@startsection{subsubsection}{3}{\z@}%
  {-0.6ex\@plus -0.2ex \@minus -.2ex}%
  {0.4ex \@plus .1ex}%
  {\normalfont\normalsize\it}}
\renewcommand\paragraph{\@startsection{paragraph}{4}{\z@}%
  {0.2ex \@plus0.2ex \@minus0.1ex}{-0.5em}%
  {\normalfont\normalsize\bfseries}}
  {\endlist}
  {\end{list}}
  {\end{list}}
  {\end{list}}
  {\end{list}}
\renewcommand\theequation{\arabic{section}.\arabic{equation}}
\numberwithin{equation}{section}
\def\ps@headings{%
  \let\@oddfoot\@empty
  \let\@evenfoot\@empty
  \def\@evenhead{\small\sffamily\thepage\hfil\slshape\leftmark}%
  \def\@oddhead{\small\sffamily{\slshape\rightmark}\hfil\thepage}%
      \let\@mkboth\markboth
    \def\chaptermark##1{\markboth{{\ifnum \c@secnumdepth >\m@ne
          \if@mainmatter \@chapapp\ \thechapter. \ \fi \fi ##1}}{}}%
    \def\sectionmark##1{\markright {{\ifnum \c@secnumdepth >\z@
          \thesection. \ \fi ##1}}}}

\advance\textwidth 4mm
\advance\hoffset -2mm
\advance\textheight 2mm
\advance\voffset -1mm
\advance\parskip by 2pt
\def\maketitle{\par\noindent{\bf\LARGE\sffamily\thetitle}\\[1.4ex]%
{\large\theauthor}\\[0.6ex]%
\textit{\small\thetextinfo}\\[0.6ex]%
{\small\noindent\today}\par\vglue1.4\bigskipamount}

\def\title#1{\def\thetitle{#1}}
\def\author#1{\def\theauthor{#1}}
\def\textinfo#1{\def\thetextinfo{#1}}

\def\reftitle#1{``#1''}
\def\[{\begin{equation}}
\def\]{\end{equation}}
\def\bse{\begin{subequations}}
\def\ese{\end{subequations}}

\makeatletter
\def\Integer{\mathbb{Z}}
\def\d{\mathrm{d}}

\def\e{\mathrm{e}}
\def\o{\mathrm{o}}

\def\KE{\mathrm{KE}}
\def\PE{\mathrm{PE}}
\def\sw{{\mathrm{sw}}}
\def\cn{{\mathrm{cn}}}
\def\E{E}
\def\tot{\mathrm{tot}}
\def\integral{\mathrm{integral}}
\def\p{\mathrm{p}}
\def\sech{\mathrm{sech}}

\def\~#1{\tilde{#1}}
\makeatother
\let\==\bar
\def\imp{\mathrm{imp}}
\def\mean{\mathrm{mean}}
\def\max{\mathrm{max}}
\def\min{\mathrm{min}}

\let\eref=\eqref

\advance\parskip 1pt

\let\trueparagraph=\paragraph
\def\paragraph#1{\par\smallskip\trueparagraph{\rm\textbf{#1}}}

\allowdisplaybreaks

\title{On the generation and propagation of solitary waves\\[0.2ex] in integrable and non-integrable nonlinear lattices}
\author{Guo Deng$^{*,\dag}$, Gino Biondini$^{*,\ddag}$, Surajit Sen$^*$ and Panayotis Kevrekidis$^\S$}
\textinfo{%
$^*$
Department of Physics, State University of New York, Buffalo, NY 14260, USA\\
$^\dag$
Department of Mathematics and Statistics, Macquarie University, 12 Wally's Walk L6, Sidney, Australia\\
$^\ddag$ 
Department of Mathematics, State University of New York, Buffalo, NY 14260, USA\\
$^\S$ 
Department of Mathematics and Statistics, University of Massachusetts Amherst, MA 01003, USA}

\begin{document}
\maketitle

\begin{abstract}
\noindent
We investigate the generation and propagation of solitary waves in the
context of the Hertz chain and Toda lattice,
with the aim to highlight the similarities, as well as differences between these systems.
We begin by discussing the kinetic and potential energy of a solitary wave in these systems,
and show that under certain circumstances the kinetic and potential
energy profiles in these systems (i.e., their spatial distribution)
look reasonably close to each other.
While this and other features, such as the connection between the
  amplitude
  and the total energy of the wave bear similarities between the two
  models,
  there are also notable differences, such as the width of the wave.
We then study the dynamical behavior of these systems in response to
an initial velocity impulse. For the Toda lattice, we do so by
employing the inverse scattering transform, and we obtain analytically
the ratio between the energy of the resulting solitary wave and the
energy of the impulse, as a function of the impulse velocity; we then
compare the dynamics of the Toda system to that of the Hertz system,
for which the corresponding quantities are obtained through numerical
simulations.
In the latter system, we obtain a universality in the fraction of
  the energy stored in the resulting solitary traveling wave
  irrespectively
  of the size of the impulse. This fraction turns out to only depend
  on the nonlinear exponent.
Finally, we investigate the relation between the velocity of the resulting solitary wave and the velocity of the impulse.  In particular, we provide an alternative proof for the numerical scaling rule of Hertz type systems.
\end{abstract}


\section{Introduction}

The existence of traveling solitary waves (SWs) --- namely, localized energy
pulses that travel without dispersing --- is a common property of many
nonlinear systems, which goes all the way back to the seminal work of
Kortweg and de Vries~\cite{kdv} on the equation that now bears their
name. 
In the past three decades, a significant focus of interest
  on such structures and their connections to numerous physical and
  engineering applications has emerged in the context of nonlinear
  dynamical lattices, i.e., in spatially discrete settings. This can
  be to a
  significant degree attributed to the emergence of the notion of
  ``granular crystals'', through the experimental as well as
  theoretical work
  of Nesterenko and collaborators~\cite{Nesterenko,Nesterenko1}
  that was subsequently expanded by numerous other groups, as
  summarized, e.g., in~\cite{sen:2008}. Importantly, developments
  have also continued along other veins,
  including the systematic study of heterogeneous lattice
  settings~\cite{vakakis}
  and the consideration of various other structures including
  shock waves and discrete breathers~\cite{chong18,IOP}. It is thus
  clear that this remains a booming field.

In the present work we build on this line of ongoing efforts.
More concretely, we explore the one-dimensional lattices of particles with nearest-neighbor interactions.
We consider lattices in which each particle has equal mass (i.e.,
homogeneous
ones), denoted by $m$.
We denote by $y_n$ the displacement of the $n$-th particle with respect to its equilibrium,
and by $\phi(r)$ the interaction potential between adjacent particles.
The equation of motion for the $n$-th particle is then given in
  this general formulation by
\vspace*{-0.2ex}
\[
m\ddot y_n=\phi'(y_{n+1}-y_n)-\phi'(y_n-y_{n-1}),
\label{e:EOM}
\]
where $n\in\Integer$, the dot denotes temporal derivative and $\phi'$ is the derivative of $\phi$.
For such discrete lattices --- also referred to as semi-discrete systems, or coupled oscillator chains --- it is well-known that
SWs exist in lattices with superquadratic potentials~\cite{FrieseckeWattis,Stefanov}.  Among lattices with superquadratic potentials, the Toda lattice~\cite{Toda,Toda1,Toda2,Toda3,Toda4} and Hertz chain~\cite{Nesterenko,Nesterenko1,sen:2008,vakakis,chong18,IOP} have been the subject of intense research over the past several decades, each for different reasons.
The Toda model corresponds to \eqref{e:EOM} with
\vspace*{-0.2ex}
\[
\phi(r_n)=\frac{a}{b}(\e^{-br_n}-1)+ar_n,
\]
where $a$ and $b$ are constants, $r_n=y_{n+1}-y_n$ is the relative displacement, and the additive constant was chosen so that $\phi(0)=0$.
In this case, \eqref{e:EOM} yields
\vspace*{-0.2ex}
\[
m \ddot y_n = a\,\e^{-b(y_n-y_{n-1})}-a\,\e^{-b(y_{n+1}-y_{n})}.
\label{e:toda}
\]
The Hertz model~\cite{Hertz}, on the other hand, corresponds to \eqref{e:EOM} with
\vspace*{-0.2ex}
\begin{eqnarray}
\phi(r_n)=
\begin{cases}
c(\Delta-r_n)^\alpha, &r_n\leq\Delta\,,\cr 0, &r_n>\Delta\,,\end{cases}
\label{e:hertzian}
\end{eqnarray}
where $c$ is a constant, $\Delta$ is the so-called precompression
  displacement,
  corresponding to the case where the chain may be compressed at its
  ends,
  inducing a displacement prior to the initiation of the nonlinear
  wave patterns.
In this case, \eqref{e:EOM} yields
\vspace*{-0.2ex}
\[
m \ddot y_n = \alpha\,c[(\Delta-(y_{n}-y_{n-1}))^{\alpha-1}-(\Delta-(y_{n+1}-y_{n}))^{\alpha-1}].
\label{e:hertz}
\]
It is also important to note in the latter case the dual nature
  of the nonlinearity stemming from the geometric exponent $\alpha$ of
  the elastic contact (Hertzian) interaction, as well as the piecewise
  definition
  of the relevant force reflecting the absence of force in the absence
  of contact~\cite{Nesterenko}.

The Hertz chain is important from a practical point of
view~\cite{Nesterenko2,Sinkovits,SenSinkovits,Coste,Hinch}, since it
serves as a model to characterize how the repelling force in the
contact area of two bodies is effected by the compression between
them.
For much of what will be considered
  below, we will limit consideration to the case of spherical
  contacts associated with $\alpha=5/2$ in our above notation,
  although some of our results will use $\alpha$ as a parameter
  motivated by a number of recent experimental developments
  enabling a certain tunability of the value of the relevant
  exponent~\cite{chong18,IOP}.
Analytical studies of Hertz systems are necessarily limited
due to the lack of complete integrability. This Hertzian system
  is
  rendered more difficult to tackle in a way in the case where the
  precompression
  displacement is absent $\Delta=0$, since in that case the system
  becomes
  ``highly nonlinear'' and has no linear limit from which to obtain
  perturbative results, e.g., in the spirit of~\cite{frpego}. I.e.,
  in the case with precompression, one can envision a potential Taylor
  expansion when $\Delta  \gg (y_{n}-y_{n \pm 1})$ and the usage of
  KdV or Toda results as a guiding principle for single solitary wave
  dynamics or multiple solitary wave interactions~\cite{Shen}. In
  the
  case of $\Delta=0$, this is no longer the case and to avoid resorting
to quasi-continuum
approximations~\cite{Nesterenko,Nesterenko1,pikovsky},
entirely different suites of mathematical techniques based on
Fourier iterative~\cite{pegoeng} and variational
methods~\cite{Stefanov}
have been developed recently.

It should be clear from the above discussion that on the one
  hand, this class of systems is of wide recent physical and
  engineering
  interest~\cite{Nesterenko,vakakis,chong18,IOP}, yet at the same time
  there is a limited set of mathematical tools and techniques in order
  to address such settings. It is the aim of the present work to
  enhance
  this toolbox by exploring similarities and differences to one of the
  most prototypical analytically tractable models in the mathematical
physics of nonlinear discrete systems, namely the Toda lattice.
The Toda lattice, as a completely integrable Hamiltonian system, possesses a rich mathematical structure which enables one to develop powerful analytical tools to obtain explicit solutions and study their behavior~\cite{Flashchka1,Flashchka2,Henon,Venakides,Deift}.

As mentioned above, both of these systems support the propagation of SWs.
This paper is devoted to characterizing {\it in a systematic fashion} the similarities and differences of the dynamics of SWs in these two models.
On one hand, because of the universal characters of SWs on nonlinear lattices, the explicit one-soliton solution of the Toda lattice can help to shed light on some of the properties of SWs in the Hertz chain.  On the other hand, these two models are intrinsically different from each other; therefore, studying the differences among them can help us to better understand the dynamics in each of these models.

The structure of this paper is the following.
In section~\ref{s:KEPE}
we begin by discussing the kinetic energy (KE) and potential energy (PE) of a SW in these two models.  We show how the oscillatory behavior of the KE and PE of a SW arises due to the discrete nature of the lattices and we show such behavior can be characterized by elliptic functions.
We also show that the PE and KE profiles of a SW in these systems look reasonably close to each other under certain circumstances.
In section~\ref{s:delta}
we then discuss the dynamics of these systems under a velocity impulse, for the Toda system by employing the inverse scattering transform (IST) we obtain an analytical expression of the ratio, between the energy of the resulting SW and the energy of the impulse, as a function of the impulse velocity, and we compare such ratio in the Toda system to numerical results in the Hertz system.
Finally, in section~\ref{s:vsw} we study the relation between the velocity of the impulse and that of the resulting SW in these systems,
in particular we provide an alternative proof of a scaling rule in Hertz-type system by using the Virial theorem.
We end this work with some final remarks in section~\ref{s:discussion}.

\section{Kinetic and potential energy of a SW}
\label{s:KEPE}

Recall that the KE and PE associated with any given system configuration are given by
\[
\label{e:KEPEdef}
\KE = \frac12\sum_{n\in\Integer}m \dot y_n^2\,,
\qquad
\PE = \sum_{n\in\Integer}\phi(y_{n+1}-y_n)\,.
\]
In both the Hertz system and the Toda lattices,
the total KE and PE of the system oscillate in time as the SW
propagates through the lattice.
This naturally reflects the shift translational invariance of the
  lattice and how the traveling wave ``reshuffles'' itself from a
  configuration centered on a lattice site to one centered between
  two lattice sites and back to the one centered on the next site, as
  it
  traverses the discrete substrate.
We next show  this oscillatory behavio due to the discrete nature
of the lattice
in a quantitative fashion. 

\subsection{SWs in the Toda lattice}
\label{ss:sw toda}
\begin{figure}[t!]
\centerline{
\includegraphics[width=0.45\textwidth]{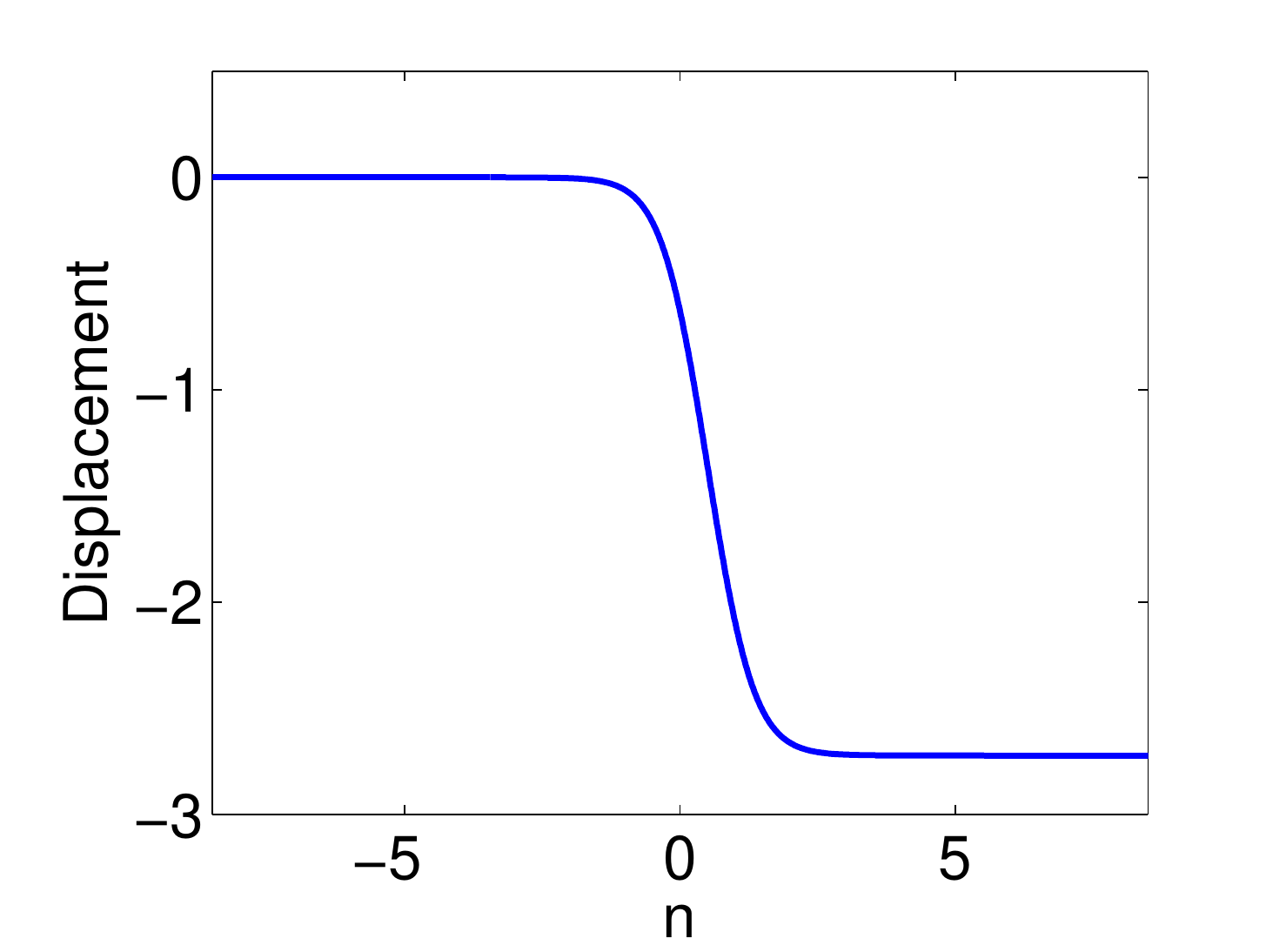}
\includegraphics[width=0.45\textwidth]{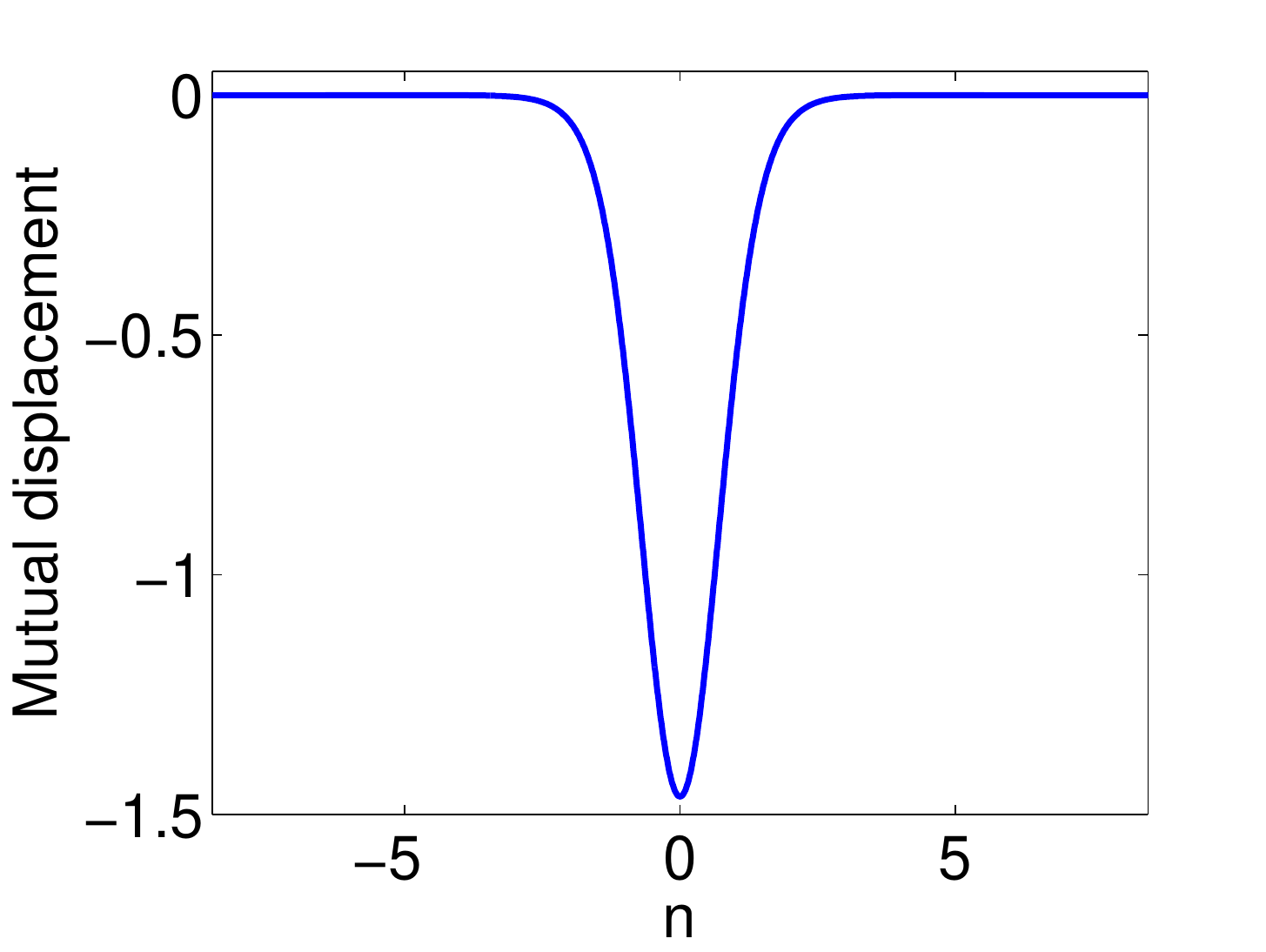}}
\caption{Left: The displacement profile $y_n(t=0)$ for a SW of the Toda lattice with $\kappa = 1.3606$.
Right: The relative displacement profile $r_n = y_{n+1}-y_n$ for the SW in the left panel.
}
\label{f:displacement}
\end{figure}

Recall that, in both systems, the displacement profile of a SW has a
kink shape.
In this work, we use the term ``profile'' to refer to the spatial distribution of certain physical quantity at fixed time $t$.
In the Toda lattice, SWs correspond to the pure one-soliton solutions of the system,
for which the displacement is explicitly given by~\cite{Toda}
\vspace*{-0.4ex}
\[
y_n(t) = \log\frac{1+\e^{2\kappa(n-1 - v_\sw t)}}{1+\e^{2\kappa(n - v_\sw t)}}+\mathrm{const},
\label{e:y_n 1 soliton}
\]
where the soliton parameter $\kappa>0$ parametrizes the family of
one-soliton solutions and for simplicity, $a$, $b$ and $m$ were set to~1.
The velocity of the SW is
\vspace*{-0.6ex}
\[
v_\sw = -\sigma (\sinh\kappa)/\kappa\,.
\]
The sign $\sigma = \pm1$ determines the direction of propagation.
A plot of the displacement profile at $t=0$ is shown in
Fig.~\ref{f:displacement}.
The plot also shows the relative displacement of the pulse,
  defined
  as $r_n=y_n-y_{n-1}$, which has the form of a pulse.
The KE and PE carried by the above SW of the Toda lattice are given by
\vspace*{-0.6ex}
\begin{subequations}
\label{e:KEPE}
\begin{gather}
\KE = \frac{\sinh^2\kappa}{2}\sum_{n\in\Integer}[\tanh(\kappa(n-1)+\sigma\sinh\kappa\, t)-\tanh(\kappa n+\sigma\sinh\kappa\, t)]^2,\\
\PE = \sum_{n\in\Integer}[\sinh^2\kappa\,\sech^2(\kappa n+\sigma\sinh\kappa\, t)]+\sum_{n\in\Integer}(y_{n+1}-y_n),
\label{e:potential}
\end{gather}
\end{subequations}
respectively.
The total energy, i.e., the sum of the KE and PE, is a constant, namely
\vspace*{-0.4ex}
\[
\E_\tot=2(\sinh\kappa\cosh\kappa-\kappa).
\label{e:solitonenergy}
\]
(To see this, note that the sum of the two infinite series in \eqref{e:KEPE} can be written as a telescoping series.)

\begin{figure}[t!]
\centerline{
\includegraphics[width=0.45\textwidth]{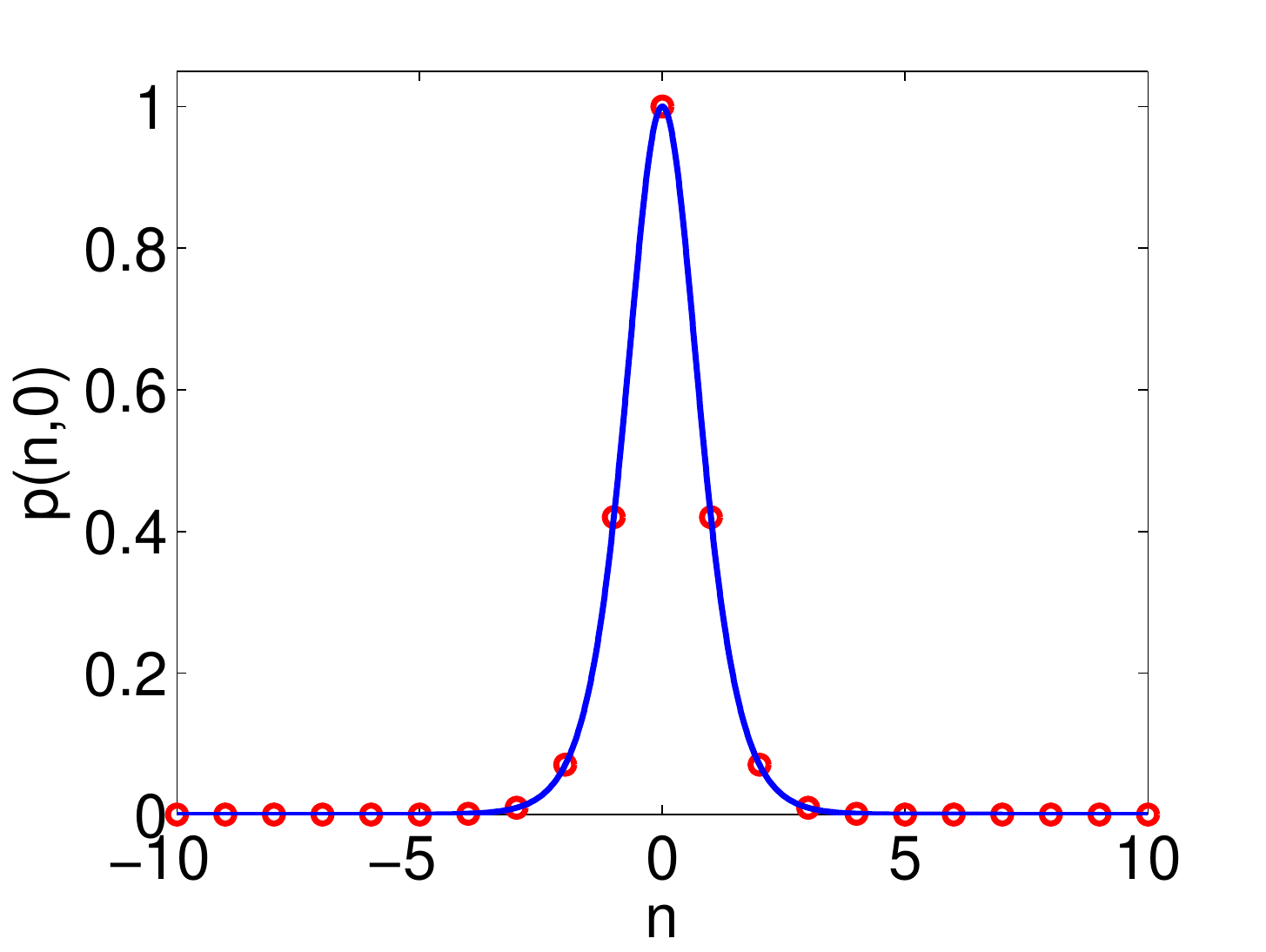}
\includegraphics[width=0.45\textwidth]{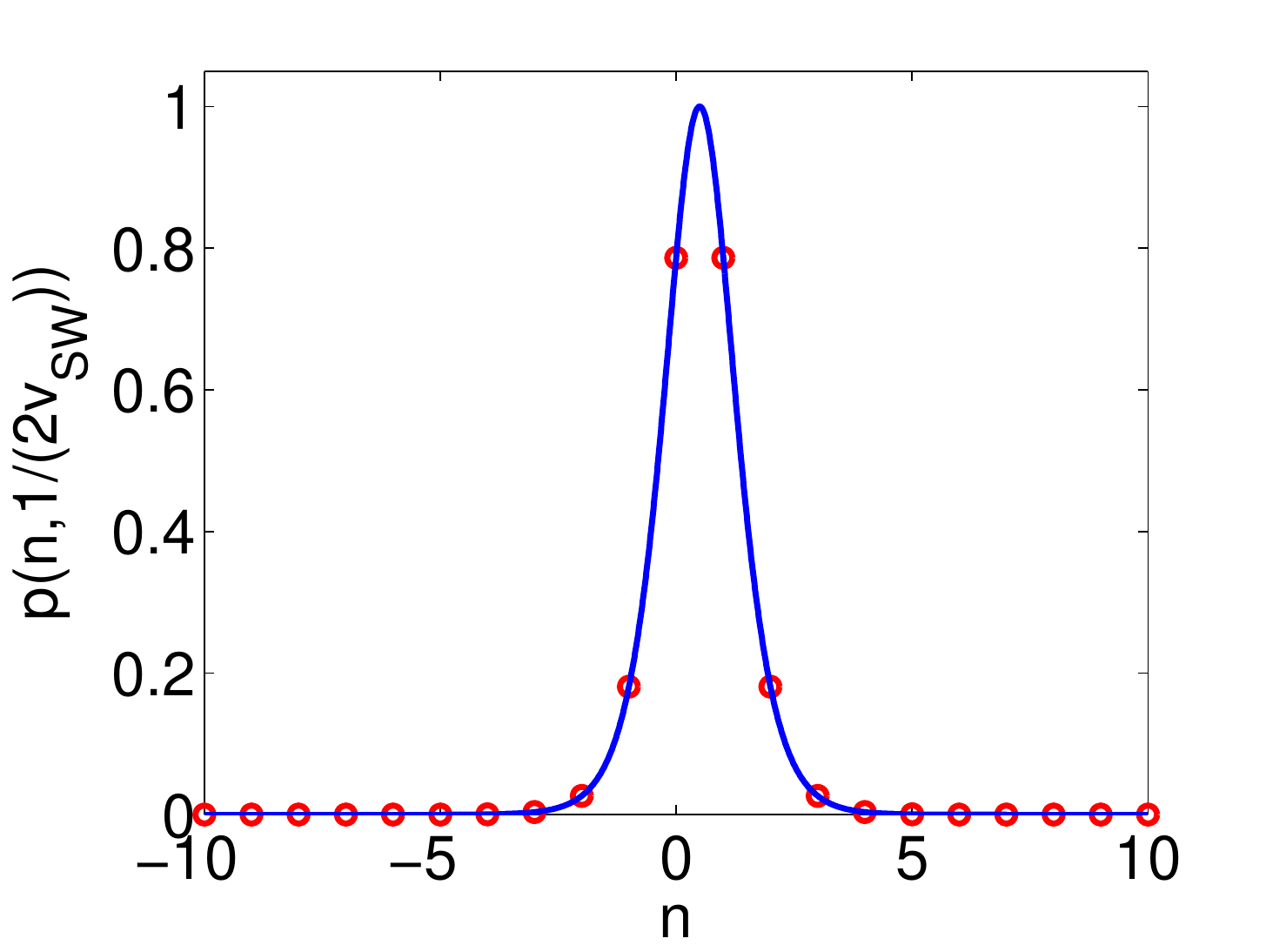}}
\caption{Left panel: the function $p(n,t)$ for the Toda lattice as a function of $n$ at $t=0$.
Right panel: $p(n,t)$ as a function of $n$ at $t=1/(2T)$.
Note that the peak of $p(n,0)$ is at $n=0$, and $\PE$ goes to maximum;
right panel: at $t=1/(2T)$, the peak of $\p(n,t)$ is between $n=0$ and $n=1$, and $\PE$ goes to minimum.
}
\label{f:pn}
\end{figure}

\begin{figure}[t!]
\centerline{
\includegraphics[width=0.405\textwidth]{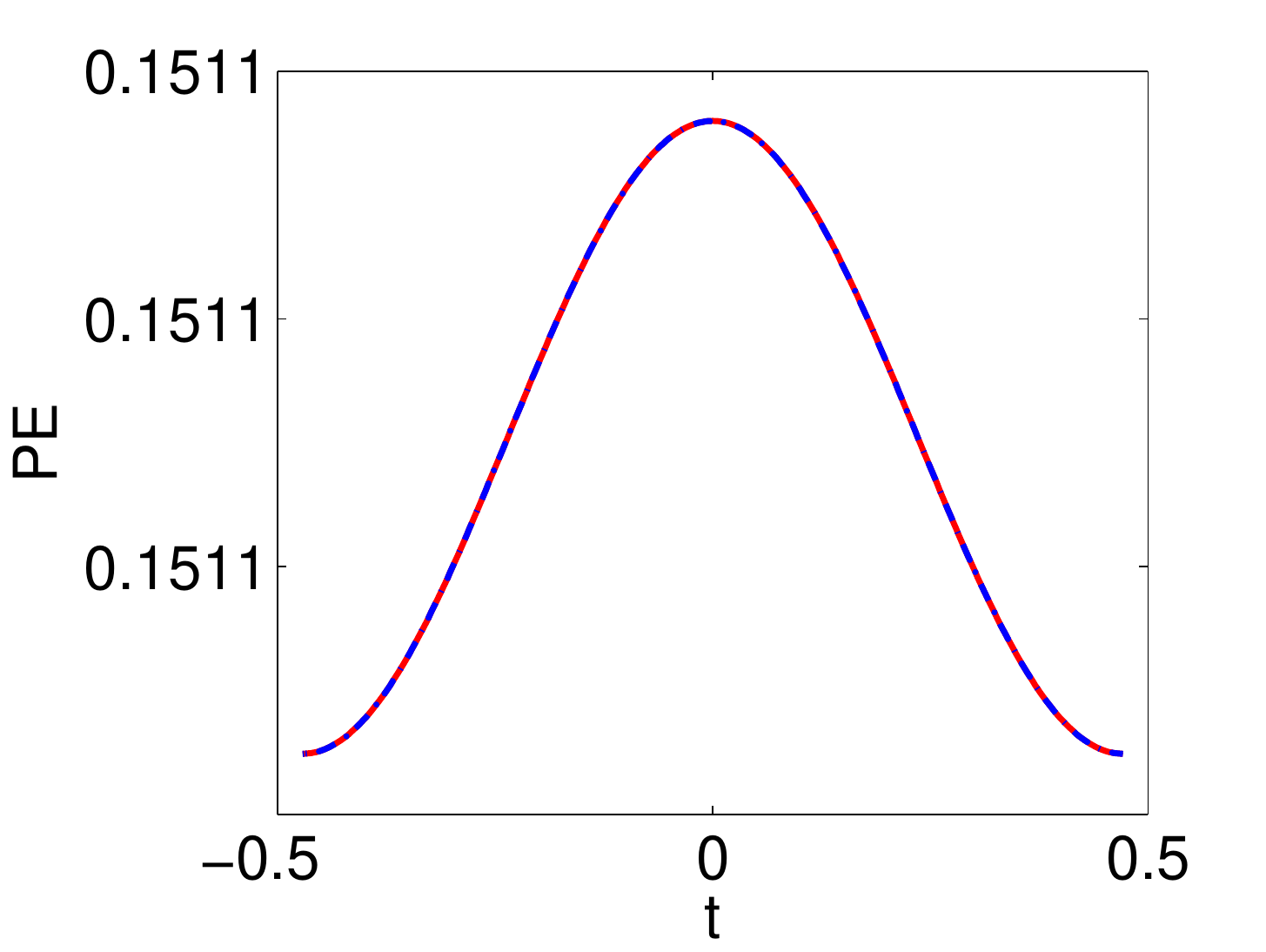}%
\includegraphics[width=0.405\textwidth]{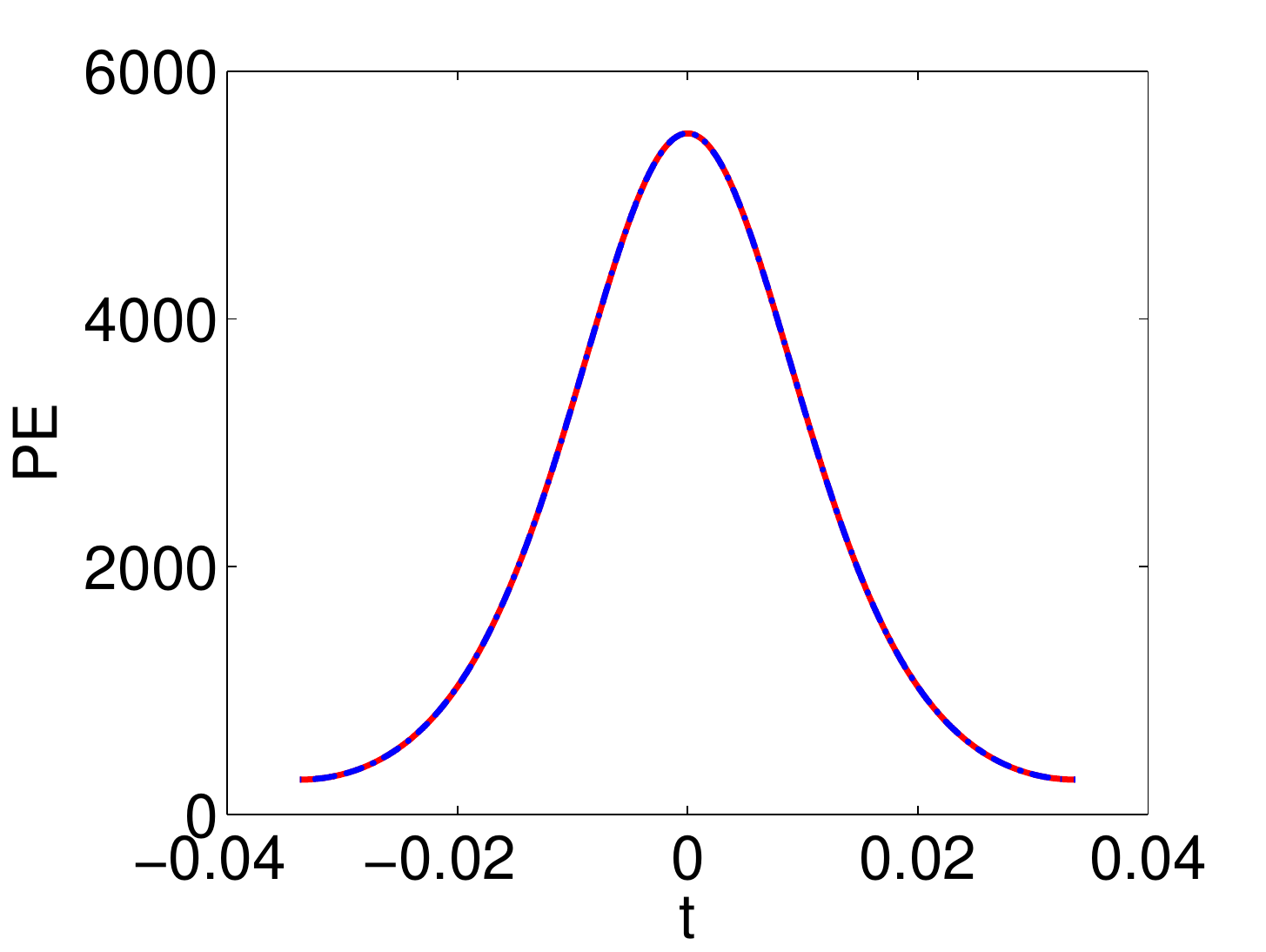}}
\caption{%
The PE for a SW in the Toda lattice as a function of time (red solid line) for a few values of the soliton parameter $\kappa$,
together with the cnoidal fitting function~\eqref{e:ellipticfit} (blue
dotted line).
One can clearly observe the exact nature of the analytical result
  which is perfectly overlapping with the numerical one.
Left: $\kappa=0.6$.
Right: $\kappa=5$.}
\label{f:oscillate}
\bigskip
\centerline{
\lower0.4ex\hbox{\includegraphics[width=0.4205\textwidth]{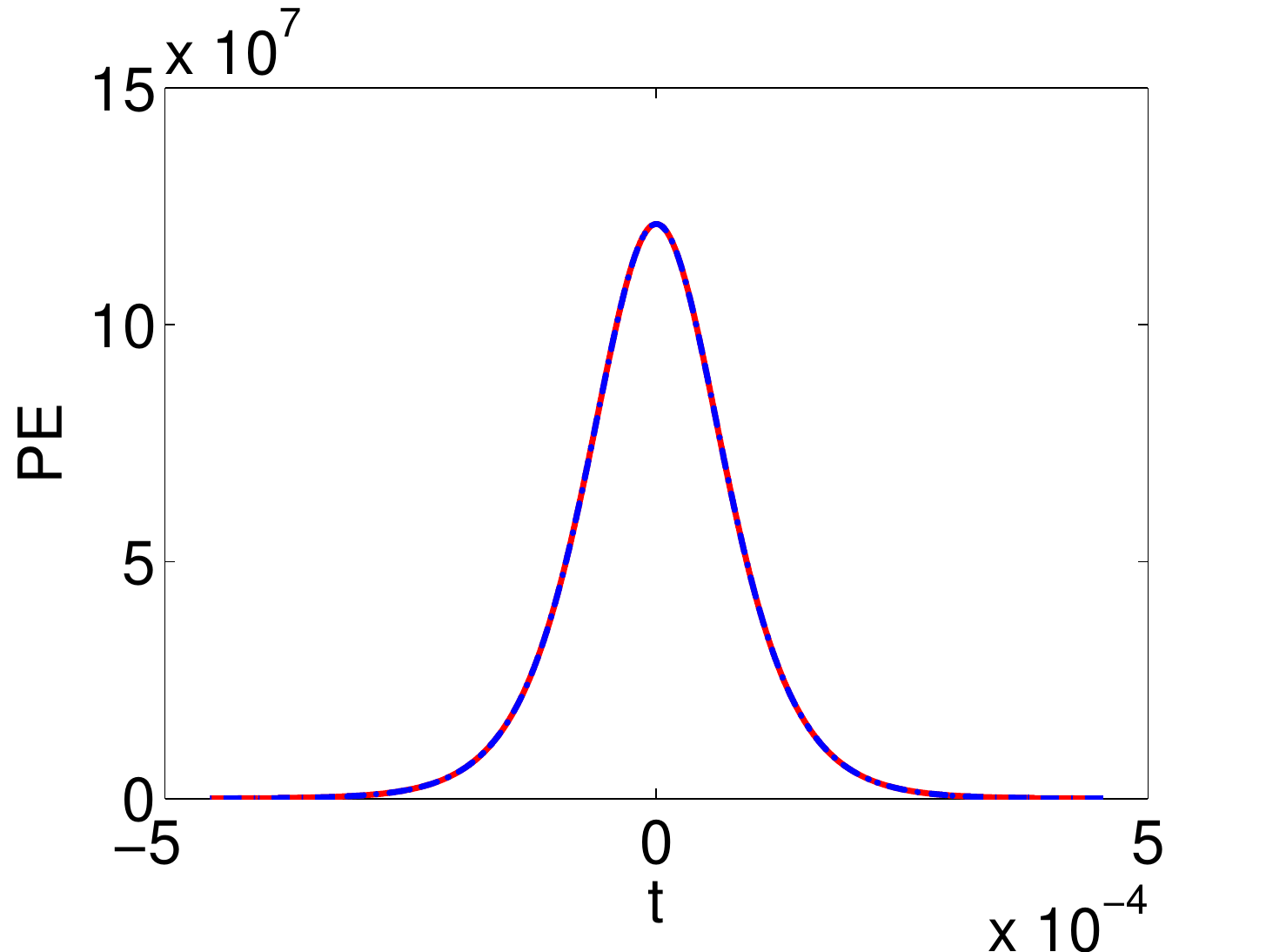}}
\includegraphics[width=0.405\textwidth]{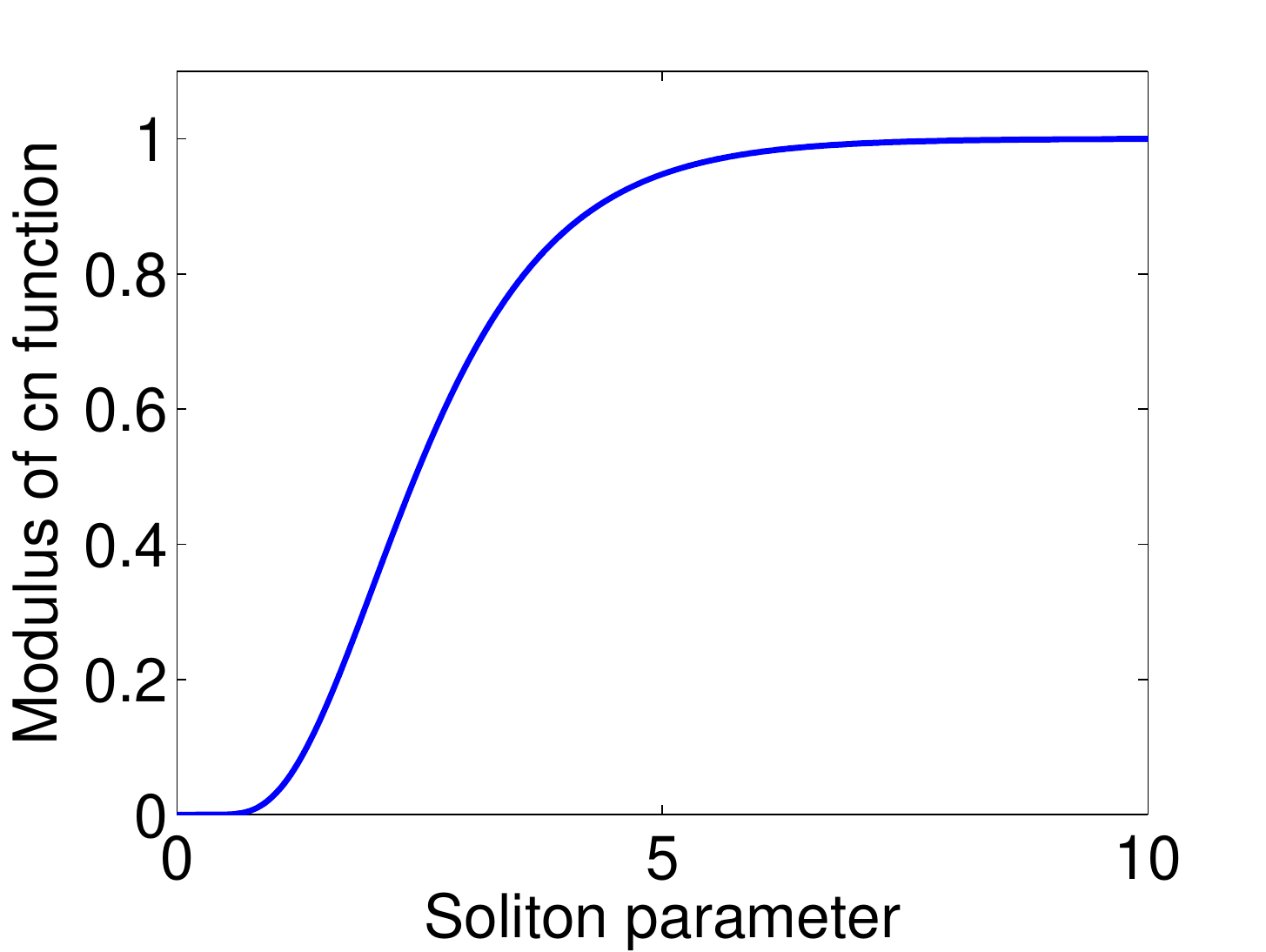}}
\caption{
Left: Same as Fig.~\ref{f:oscillate}, but for $\kappa = 10$.
Right: The elliptic parameter $m$ in~\eqref{e:sech_cn} as a function of the soliton parameter~$\kappa$.}
\label{f:diffmod}
\end{figure}

The discrete nature of lattice induces an oscillatory behavior in the
KE and PE.
First off, notice how the pulse in the relative displacement
  alternates between an onsite and an intersite spatial structure
  in Fig.~\ref{f:pn}, as discussed above.
Figure~\ref{f:oscillate} shows the temporal oscillations of the PE for a few different values of the soliton parameter~$\kappa$.
At $t=0$, the peak of $p(n,t)$ as a function of $n$ overlaps with the particle placed at $n=0$, as shown in the left panel of Fig.~\ref{f:pn}.
Moreover, the sum in~\eqref{e:PEmod} reaches its maximum value as function of time at $t=0$.
Conversely, at $t= 1/(2v_\sw)$
the peak of $\p(n,t)$ lies at the midpoint between the particles located at $n=0$ and $n=1$,
as shown in the right panel of Fig.~\ref{f:pn}.
At this point in time, the sum in~\eqref{e:PEmod} reaches its minimum value.
Finally, at $t= 1/v_\sw$, the peak of $\p(n,t)$ overlaps with the particle at $n=1$ and the sum goes back to its maximum value.
Therefore $\PE$ and $\KE$ oscillate periodically in time, with period given by $T = 1/|v_\sw|$.

Note that the first term in~\eref{e:potential} depends on time, while the second term is constant and equals $-2\kappa$.
Therefore, it is convenient to express the PE as
\[
\PE(t) = -2\kappa + \sinh^2\kappa\,\sum_{n\in\Integer}p(n,t),\qquad
p(n,t) = \sech^2(\kappa n+\sigma\sinh\kappa\, t).
\label{e:PEmod}
\]
We now make use of the identity~\cite{NIST}
\begin{equation}
\sum_{n=-\infty}^{\infty}\bigg[\sech^2\bigg(\frac{\pi}{2K'_m}(z - 2K_m n )\bigg)\bigg]
  =\bigg(\frac{2K'_m}{\pi}\bigg)^2\bigg[\bigg(\frac{E'_m}{K'_m} - m \bigg)+ m\,\cn^2(z;m)\bigg],
\label{e:sech_cn}
\end{equation}
where $\cn(\cdot)$ is one of the Jacobian elliptic functions,
$m$ is the elliptic parameter (i.e., the square of the elliptic modulus),
$K_m = K(m)$ and $E_m = E(m)$ are the complete elliptic integral of the first and second kind respectively,
$K'_m = K(\sqrt{1-m})$ and $E'_m = E(\sqrt{1-m})$.
Thus, the oscillatory behavior of the PE is characterized by elliptic oscillations.
Indeed, comparing~\eref{e:PEmod} and~\eref{e:sech_cn}, we obtain the PE as a function of $t$ is given by
\[
\PE(t) = \PE_{\min}+\Delta\PE\,\cn^2(\Omega t;m),
\label{e:ellipticfit}
\]
where
\begin{subequations}
\label{e:deltape}
\begin{gather}
\PE_{\max} = \sinh^2\kappa\bigg(\frac{2K'_m}{\pi}\bigg)^2\frac{E'_m}{K'_m}-2\kappa\,,
\qquad
\PE_{\min} = \sinh^2\kappa\bigg(\frac{2K'_m}{\pi}\bigg)^2\bigg(\frac{E'_m}{K'_m}-m\bigg)-2\kappa\,,
\\
\noalign{\noindent the oscillation amplitude is}
\Delta\PE = \PE_{\max}-\PE_{\min} = \sinh^2\kappa
\bigg(\frac{2K'_m}{\pi}\bigg)^2 m\,,
\end{gather}
\end{subequations}
and the elliptic parameter $m$ is implicitly determined by the relation
\[
\kappa=\pi K_m/K'_m.
\label{e:kappam}
\]
The value of $m$ as a function of $\kappa$ is shown in Fig.~\ref{f:diffmod},
The oscillation frequency is $\Omega = 2K_m/(\kappa/\sinh\kappa)$, and the oscillation period is therefore given by $T=\kappa/\sinh\kappa\equiv1/v_\sw$.
Note that the PE attain the maximum at $t=0,T,2T\dots$, and attain the minimum at $t=T/2,3T/2\dots$.
It is worthwhile to note here that similar calculations have been
  previously performed in the context of theories bearing SWs
  (including
  integrable ones, such as the so-called Ablowitz-Ladik lattice)
  in the work of~\cite{cai}. There, the principal technique used
  involves the Poisson summation formula. Yet, given the above
  mentioned identity of Eq.~(\ref{e:sech_cn}),
  we do not follow such a methodology here.

\subsection{SWs in the Hertz system and comparison}

No analytical expression is known for the shape of SWs in the Hertz system.
Here, we therefore discuss the similarities and differences between SWs in the Toda system and the Hertz system.

The displacement profile of SWs in the Hertz system is also kink-shaped, like that in the Toda system.
An expression for the SWs in the Hertz system without precompression (i.e., with $\Delta=0$) was given in~\cite{SenManciu} as
\vspace*{-0.4ex}
\[
\label{e:sw}
y_n(t) = Y(n - v_\sw t)\,,
\quad
Y(z)=\frac{A}{2}\bigg\{1-\tanh\bigg[\frac{f(z)}{2}\bigg]\bigg\},\quad
 f(z)=\sum_{q=0}^\infty C_{2q+1}z^{2q+1},
\]
where $A$ denotes the SW amplitude,
and the constants $C_m$ are obtained numerically via an iterative algorithm.
For $\alpha = 2.5$, one finds
$C_1=2.39536$, $C_3=0.268529$, $C_5=0.0061347$ \cite{SenManciu}.
It should be noted that this is not the only representation of
  this approximate kind of the SW. An alternative one has been
  offered on the basis of Pad{\'e} approximations, e.g.,
  in~\cite{yuli}.
  On the other hand, the exact numerical form (up to a controllable
  numerical
  tolerance) of the SW on the basis
of an iterative Fourier technique was identified in~\cite{pegoeng}.

\begin{figure}[t!]
\centerline{
\includegraphics[width=0.45\textwidth]{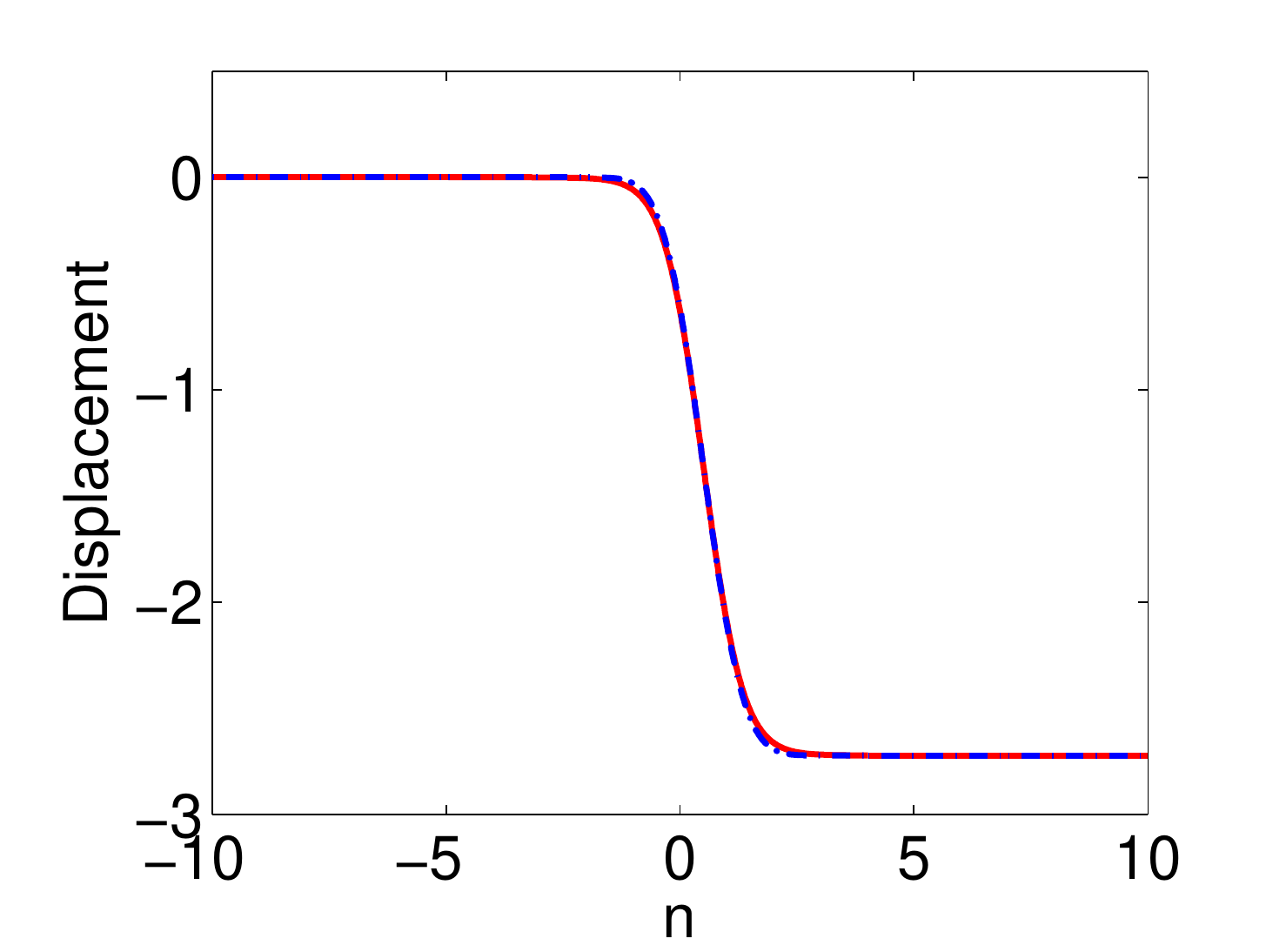}
\includegraphics[width=0.45\textwidth]{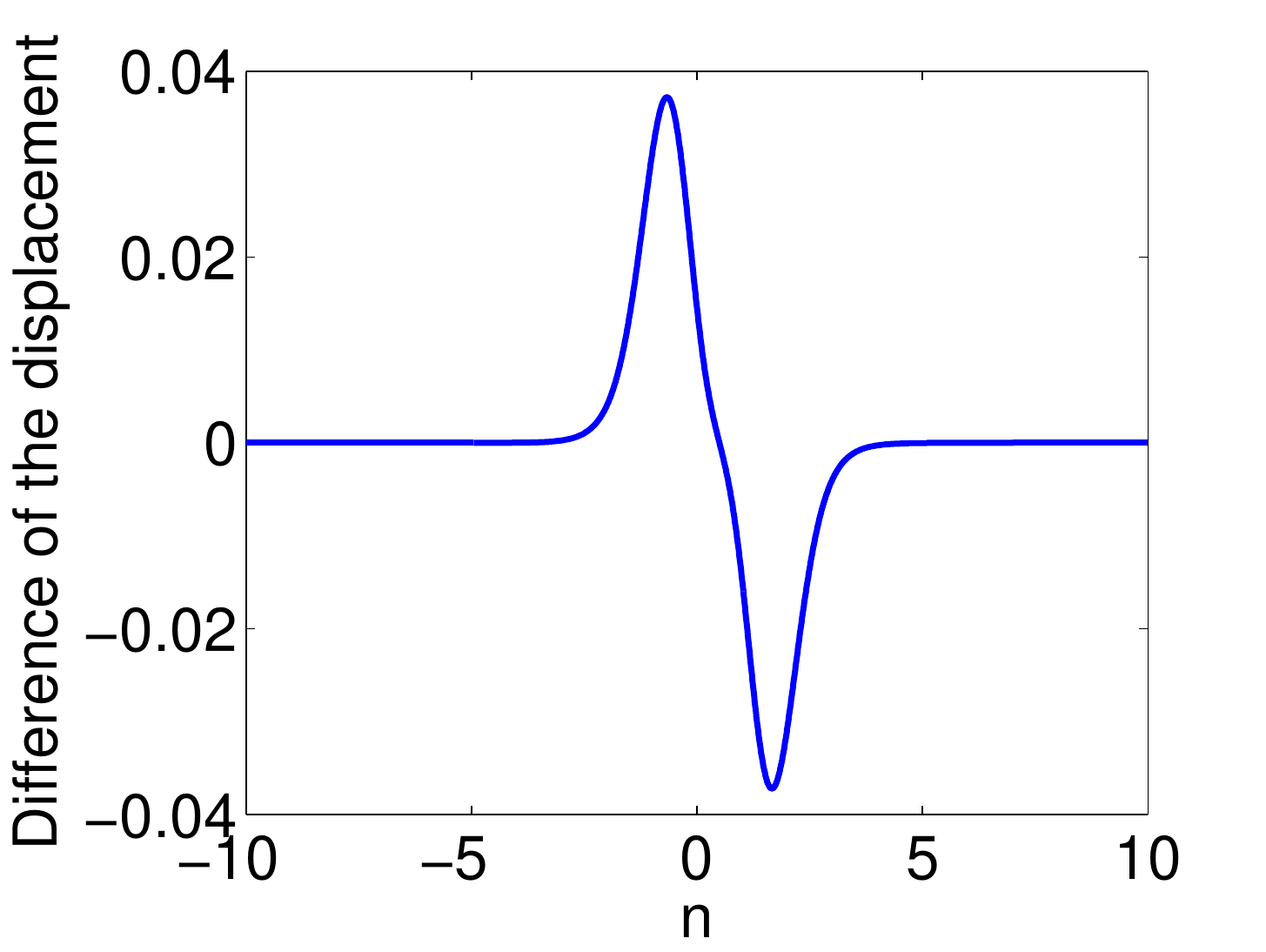}}
\caption{Left: Red line represents the displacement profile for a SW of the Toda lattice with $\kappa = 1.3606$ at $t=0$ and the blue dotted line represents the displacement profile for a SW of the Hertz chain with the same amplitude when $\alpha=2.5$.
Right: Difference between the two curves in the left panel.
}
\label{f:displacement2}
\centerline{\includegraphics[width=0.45\textwidth]{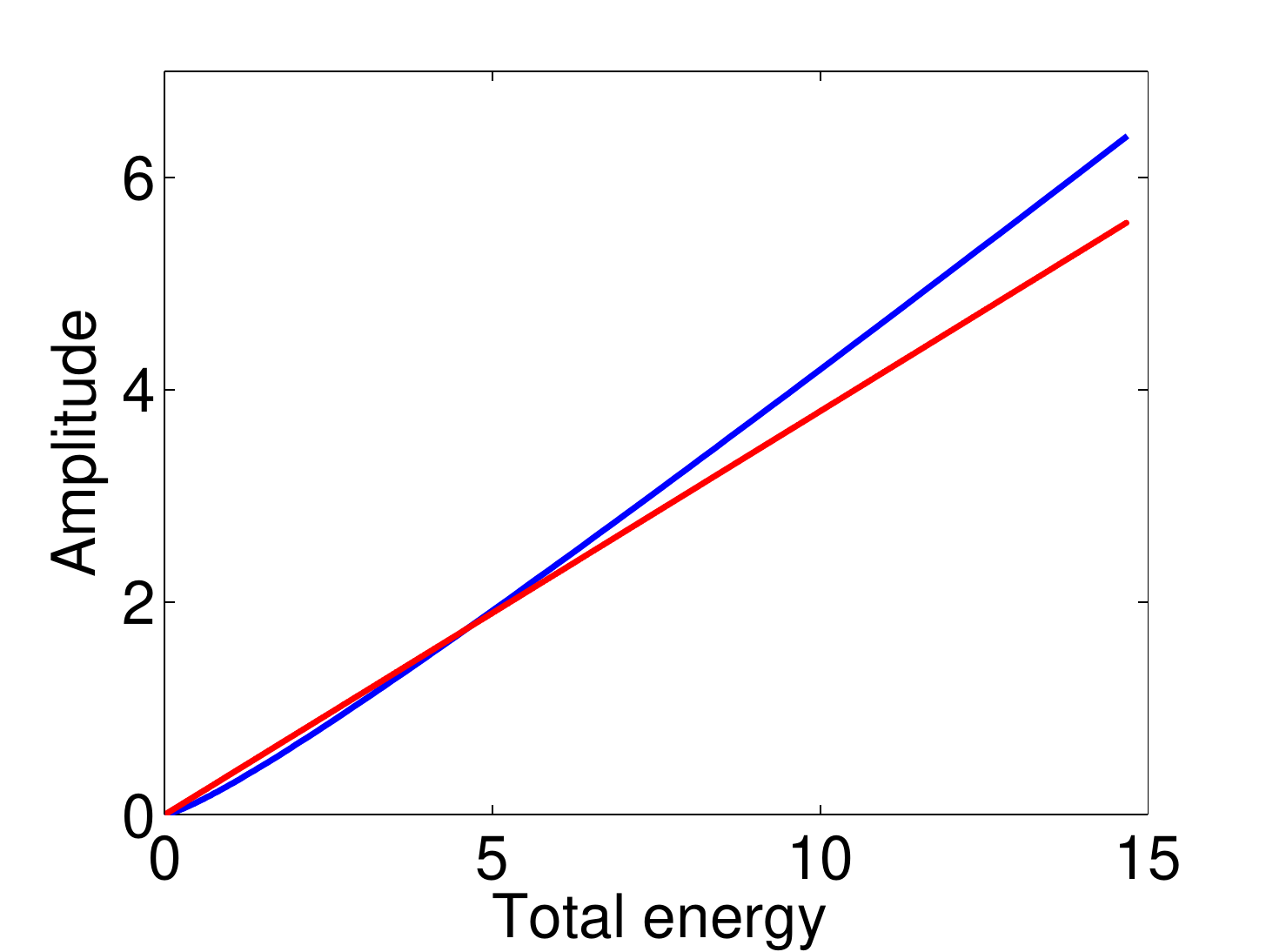}
\includegraphics[width=0.45\textwidth]{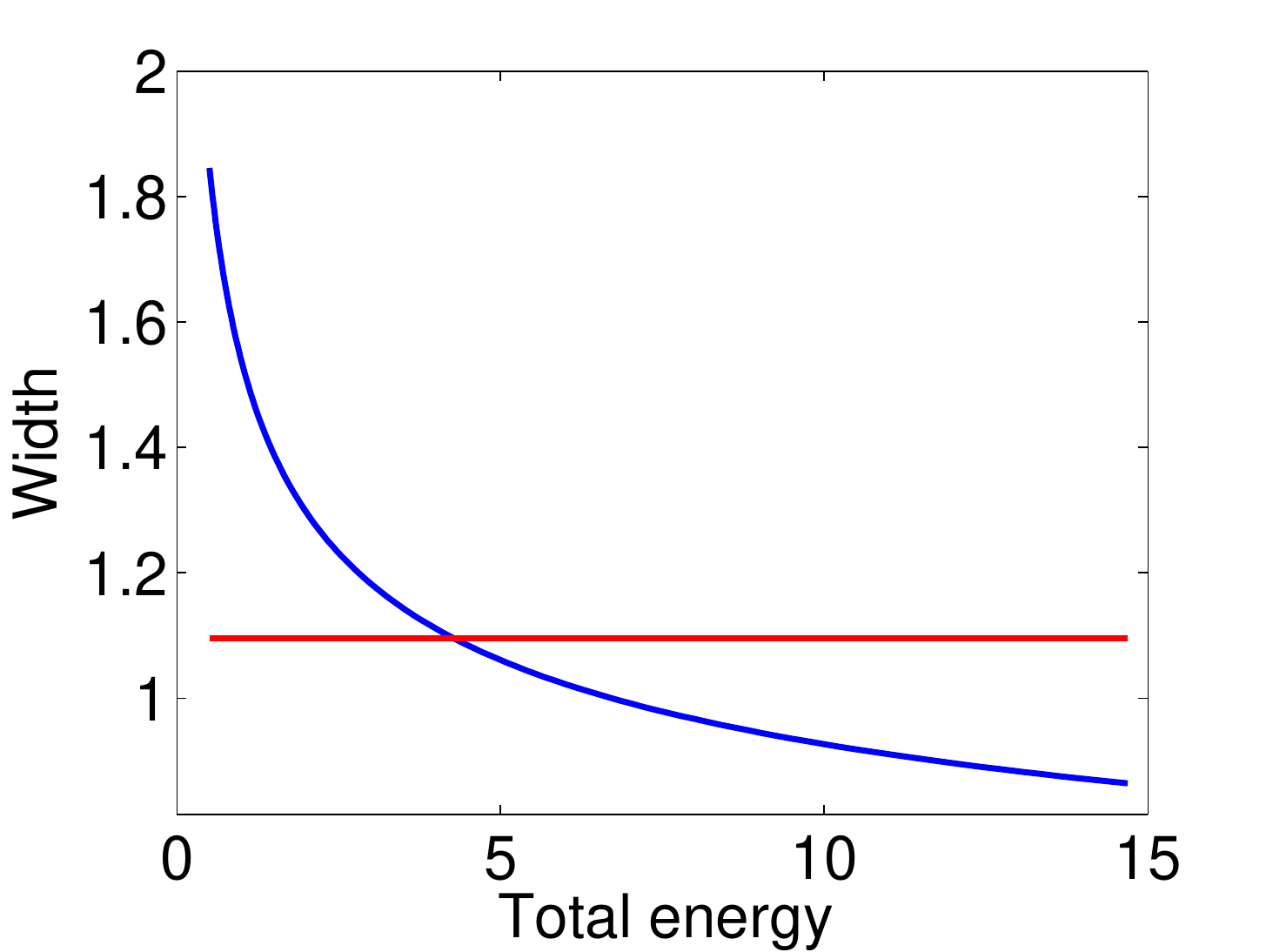}}
\caption{Left: Amplitude of the PE profile for a SW in the Toda system (blue line) and the Hertz system with $\alpha=2.5$ (red line) as a function of energy.
Right: Width of the PE profile for a SW in the Toda system (blue line) and the Hertz system (red line) as a function of energy.
}
\label{f:hertziantoda}
\end{figure}

The overall shape of SWs in the two systems is qualitatively very similar,
as shown in Fig.~\ref{f:displacement2}. 
                 Nonetheless, there are important differences.
In the Toda system, SWs with higher energy are narrower.
In contrast,
the width of SWs in the Hertz system is independent of the energy of the SW, as shown in Fig.~\ref{f:hertziantoda}.
This is because the coefficients $C_m$ --- and therefore $f(z)$ --- are independent of the SW energy.
Thus, in this sense, in the Hertz system the shape of SWs is universal
{and only the amplitude (and speed) scales, while the width remain
  unchanged}.
Moreover, Fig.~\ref{f:hertziantoda} also shows that the amplitude of the PE profile is linearly proportional to energy.
A direct consequence of~\eqref{e:sw} is that the PE of SWs in the Hertz system scales like $A^\alpha$.
Thus, we immediately obtain that the total energy of a SW in the Hertz system also scales like $A^\alpha$.

\begin{figure}[t!]
\centerline{\includegraphics[width=0.33\textwidth]{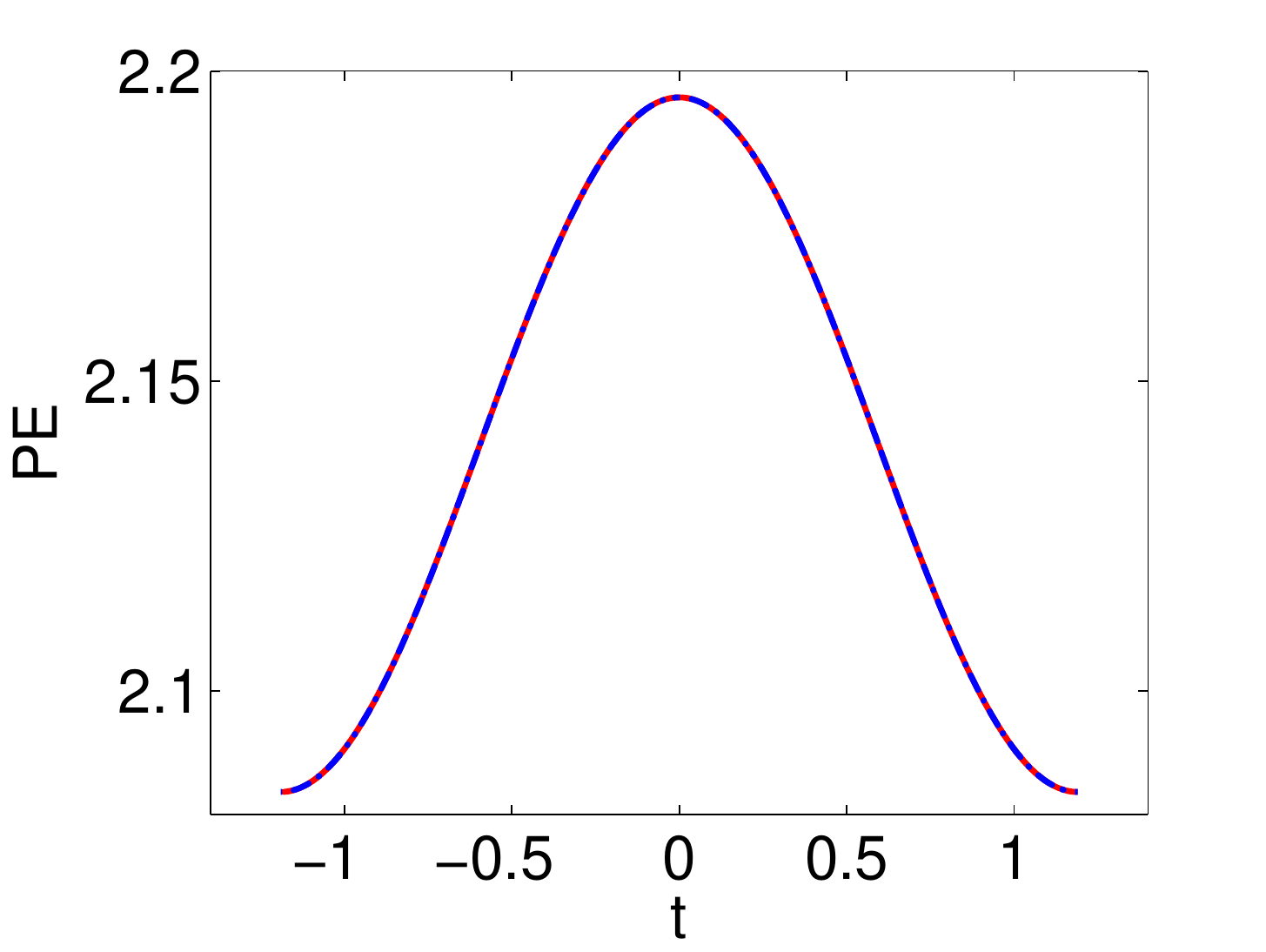}
\includegraphics[width=0.33\textwidth]{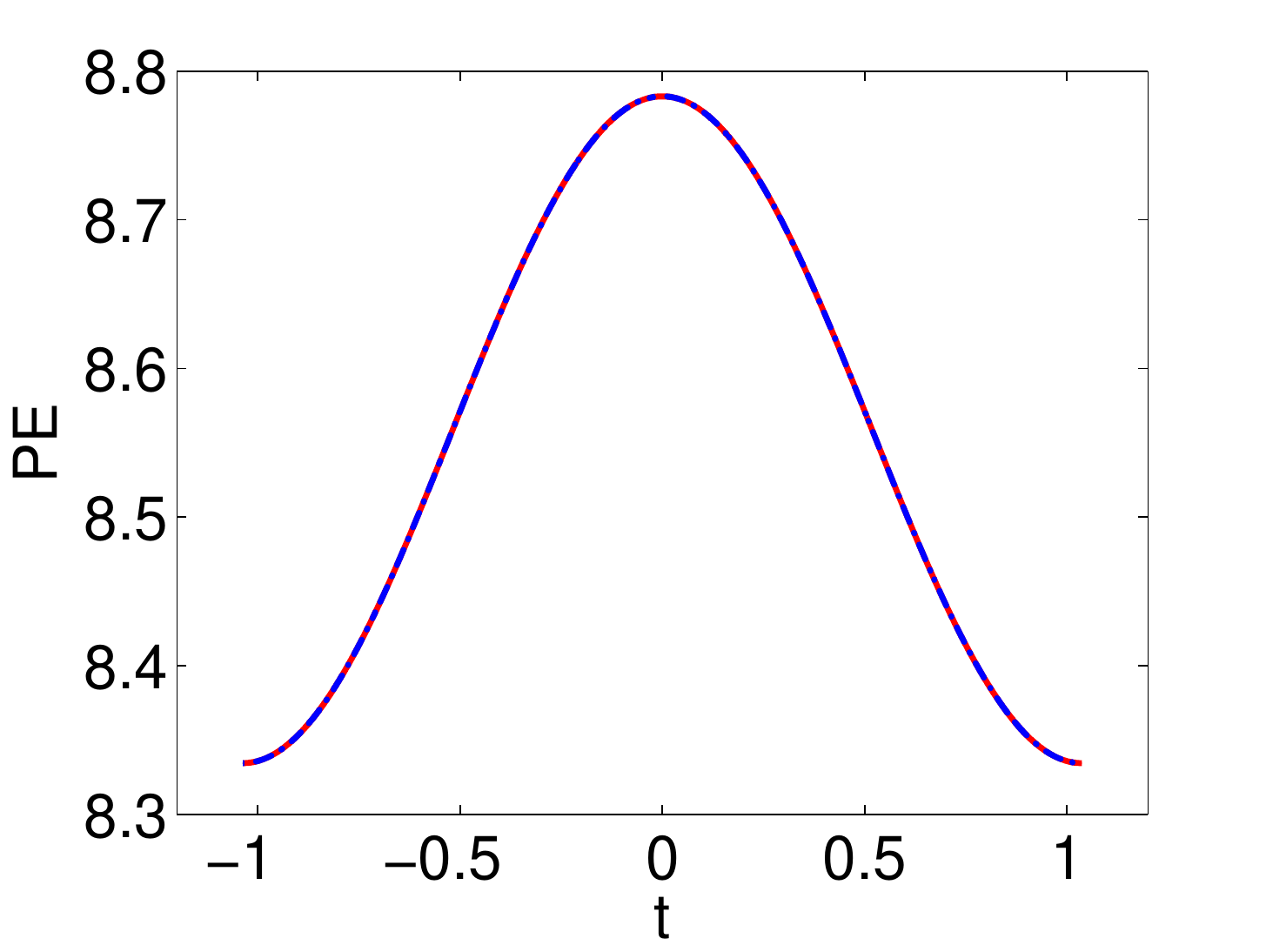}
   \lower-0.2ex\hbox{\includegraphics[width=0.33\textwidth]{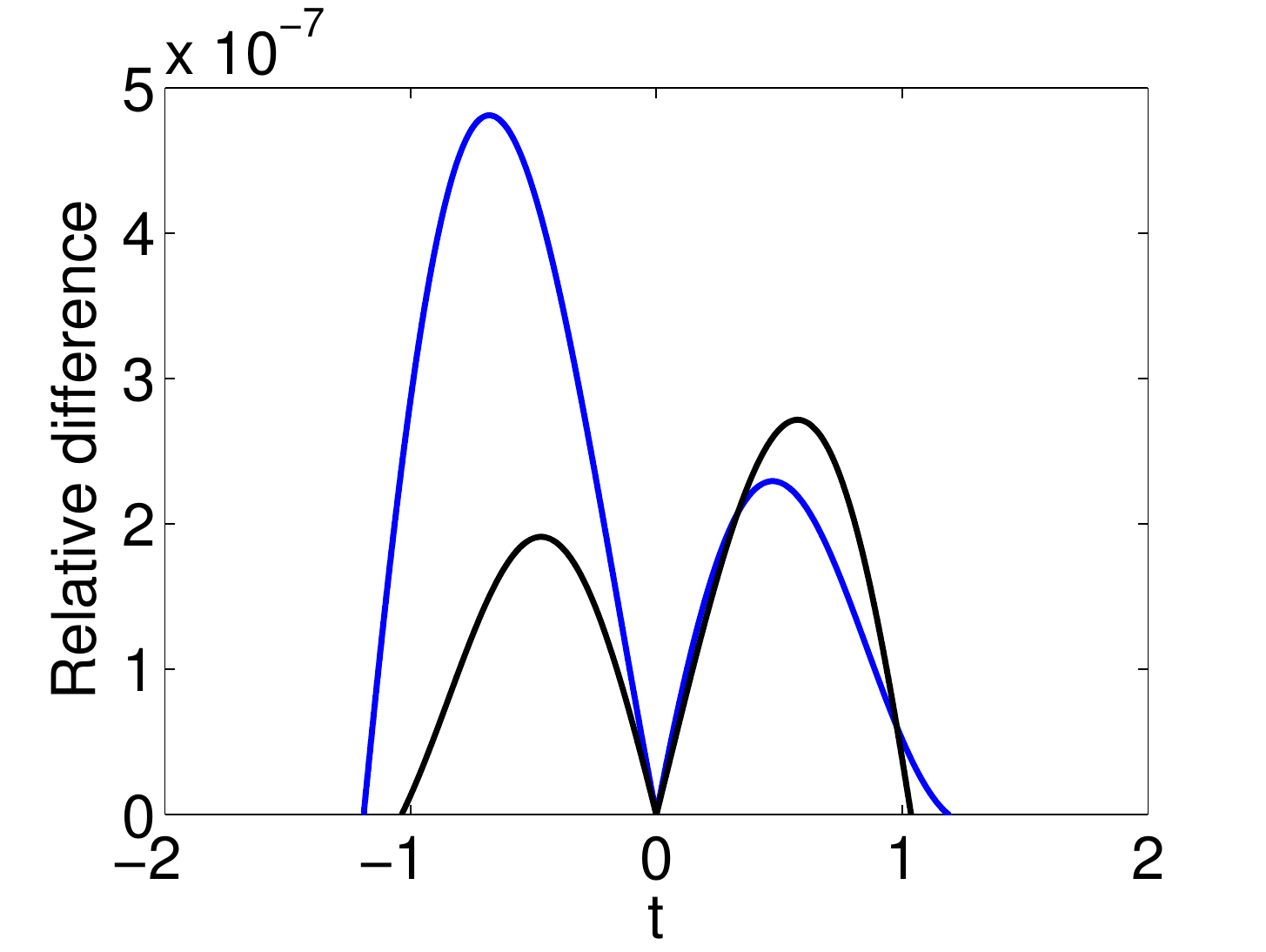}}
         }
\caption{
Left panel and middle panel: The PE (red curve) of a SW for the Hertz
chain with $\alpha=2.5$ as a function of time at $E=4.83$ and $E=9.66$
respectively. The blue dotted line represents the sinusoidal fitting
function.
It can be clearly seen that the relevant fit is excellent.
Right panel: The relative difference between the PE and the sinusoidal fitting function at $E=4.83$ (blue line) and $E=4.96$ (black line) respectively.
}
\label{f:pe_hertzian}
\end{figure}

Because of the kink-shaped displacement profile, the KE and PE profiles in the Hertz system are also bell-shaped, like those in the Toda system.
The amplitude and the full width at half maximum of the PE profile as a function of $\E$ for both systems is shown in Fig.~\ref{f:hertziantoda}.
Therefore, just like in the Toda system,
the PE and KE of a SW in the Hertz system also demonstrate oscillatory behavior.
In fact, as shown in Fig.~\ref{f:pe_hertzian},
such behavior is well characterized by sinusoidal functions at all values of energy.
(The relative difference between the PE and the fitting function is given in the right panel of Fig.~\ref{f:pe_hertzian}.)
Recall that, in the Toda system, the oscillatory behavior of the PE is instead characterized by an elliptic cn,
and the elliptic modulus~$k$ is determined by the soliton parameter~$\kappa$ which is in one-to-one correspondence with the width of the SW.
Also recall that the elliptic cn reduces to a sinusoidal function when
$k \rightarrow 0$.
Thus, one way to interpret the difference between the oscillatory behavior in the Toda system and in the Hertz system is to recall that
the width of the SW in the Hertz system is independent of energy,
and therefore so is the shape of the temporal oscillations of the PE.

There is also a significant deviation in the nature of the tails
  of the different waves which is partially reflected in the
  Fig.~\ref{f:displacement2}. The Toda soliton is exponential in its
  decay reflecting the linear ``component'' of the exponential
  interaction
  force between adjacent beads. On the other hand, the highly
  nonlinear
  (for $\Delta=0$) Hertzian model and hence the associated SW
  possess no linear component. Thus the decay of the wave is
  doubly exponential and generally considerably faster in a semilog
  plot than its Toda counterpart~\cite{pegoeng,Stefanov}.

\begin{figure}[t!]
\centerline{\includegraphics[width=0.45\textwidth]{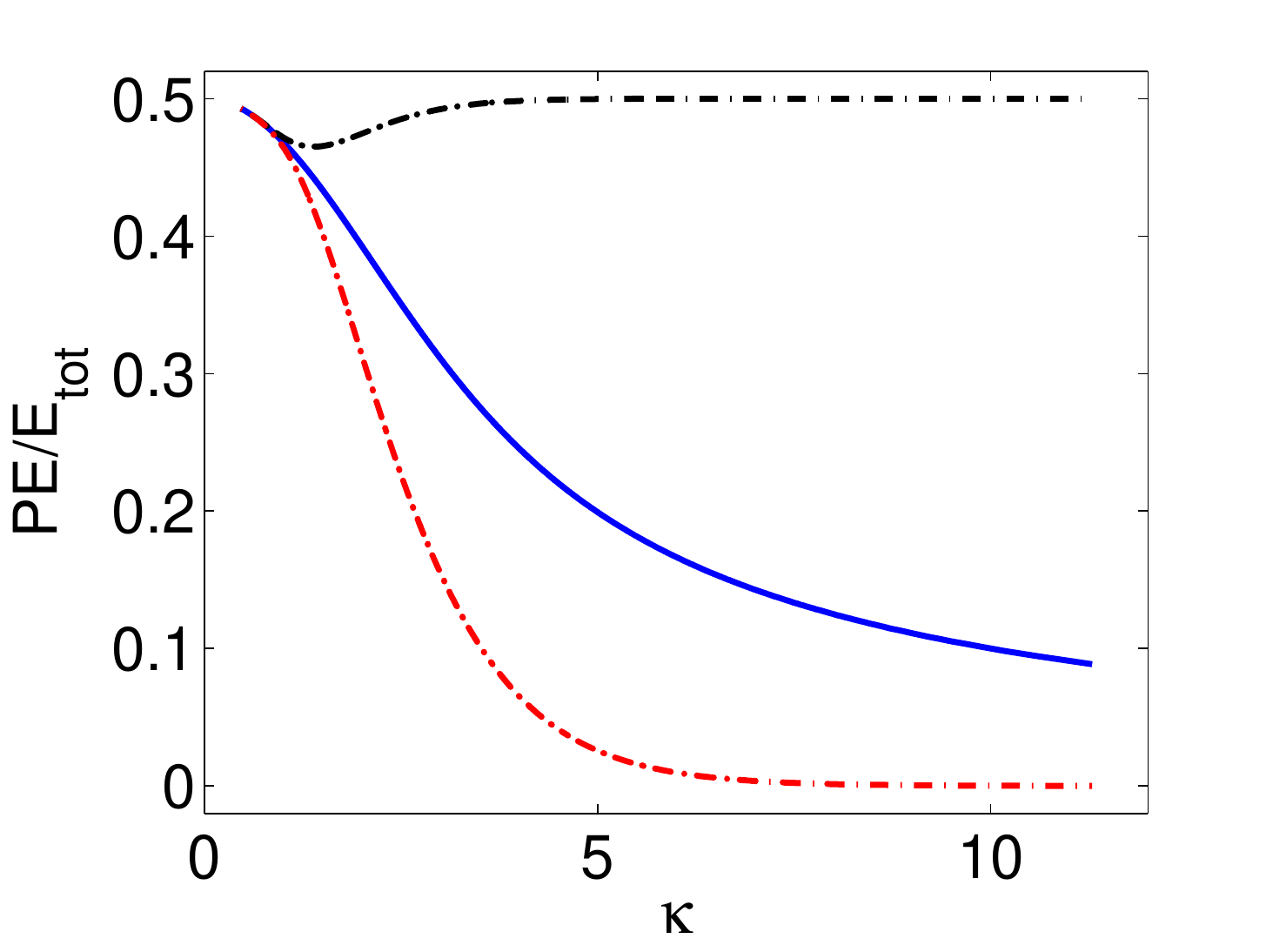}
            \includegraphics[width=0.45\textwidth]{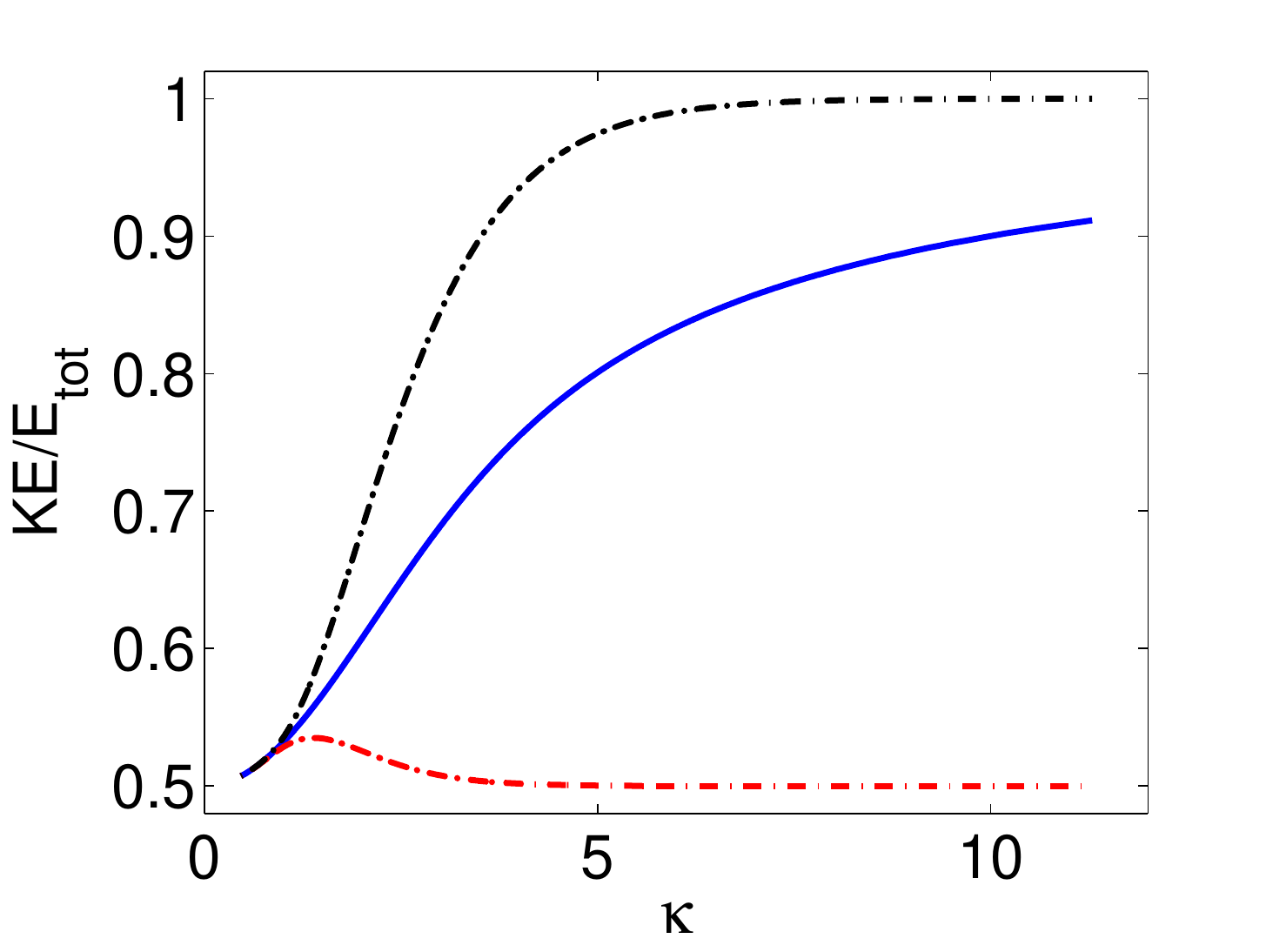}}
\caption{Left panel: The ratio between $\PE_{\min}$, $\PE_{\max}$ and $\overline{\PE}$ (dash-dotted red curve, dash-dotted black curve and solid blue curve, respectively)
and the total energy of a SW in the Toda system as a function of the soliton parameter~$\kappa$.
       Right panel: The ratio between $\KE_{\min}$, $\KE_{\max}$ and $\overline{\KE}$ and the total energy of a SW in the Toda system as a function of the soliton parameter~$\kappa$.}
\label{f:peetot}
\bigskip
\centerline{
\includegraphics[width=0.45\textwidth]{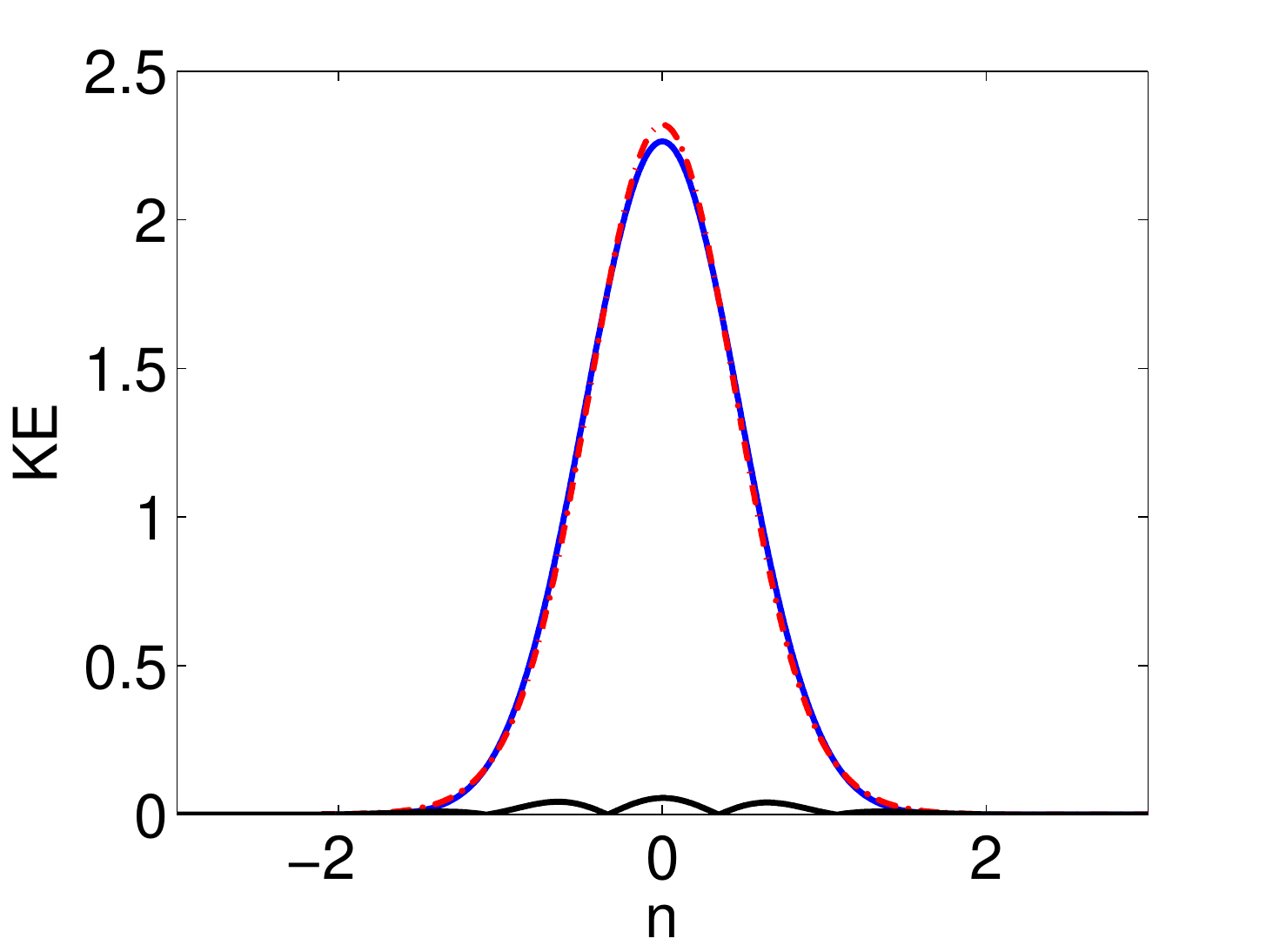}
\includegraphics[width=0.45\textwidth]{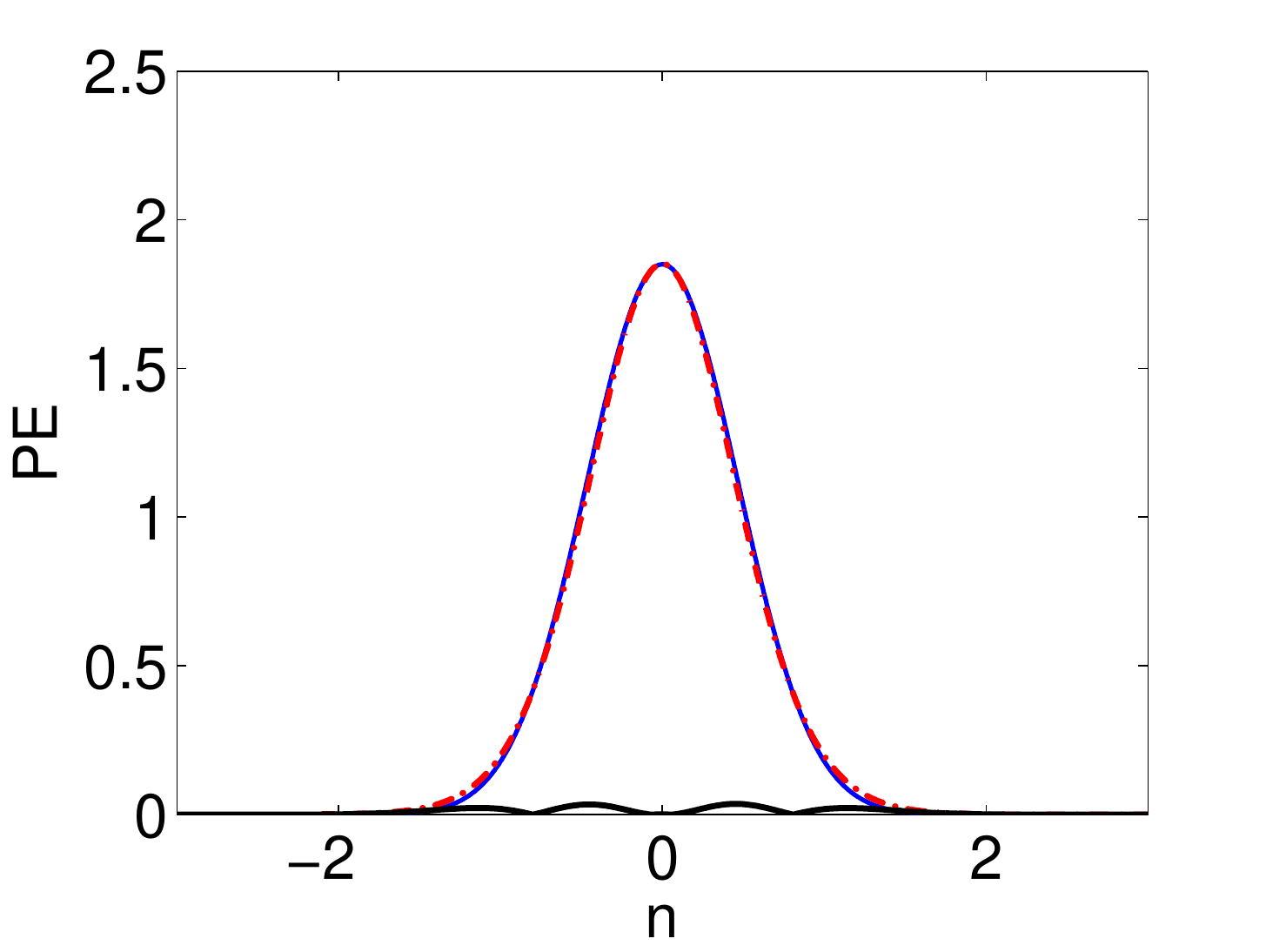}}
\caption{Left panel: The KE profile for a SW with the same fraction of kinetic and potential energy in the two systems. Right panel: The PE profile for the same SW. The red line represents the Hertz system with $\alpha=2.5$; the blue line represents the Toda system, and the black line represents the difference between them.
}
\label{f:peke}
\end{figure}

Since the properties of SWs in the Toda system and the Hertz system have a different dependence on energy,
there are several different ways to compare the two systems.
So far, we have identified three possible criteria: matching the SW energy, matching its amplitude, and matching its width.
A fourth criterion, however, is related to how a SW in each of these systems distributes the total energy $\E_\tot$ between the kinetic and potential energy.
For this purpose, it it useful to look at the value of $\overline{\PE}/\E_\tot$, where
\[
\overline{\PE}=\frac1T\int_0^T\PE(t)\d t,
\label{e:peavg}
\]
denotes the temporal average of the potential energy over a period.

For the Hertz chain we simply have $\overline{\PE}=\PE_\mean$, where $\PE_\mean =\frac12(\PE_\max + \PE_\min)$.
This is because the temporal oscillations of the PE in the Hertz chain are sinusoidal.
For the Toda lattice, however, $\overline{\PE}$ only equals $\PE_\mean$ in the limit $\kappa\to0$,
because the temporal oscillations only become sinusoidal in that limit.
On the other hand, the integral in~\eqref{e:peavg} can be computed explicitly using~\eqref{e:ellipticfit}, to obtain
\[
\overline{\PE}=\PE_{\min}+\frac{4\sinh^2\kappa}{\pi^2} K_m'^2\bigg(\frac{E_m}{K_m}-1+m\bigg)
\label{e:peavg1}
\]
for the Toda lattice.

The virial theorem~\cite{goldstein}
states that any bounded system with interaction potential $\phi(r)=r^\alpha$ satisfies the relation $2\overline{\KE}-\alpha\overline{\PE}=0$.
As a result, for the Hertz system without precompression (i.e., with $\Delta=0$), one has $\overline{\PE}/\E_\tot=4/9$ for all SWs.
The virial theorem does not apply to the Toda lattice, because the interaction potential is not simply a power law.
In that case, instead, $\overline{\PE}/\E_\tot$ decreases monotonically as the soliton parameter $\kappa$ increases as shown in Fig.~\ref{f:peetot}.
In particular, in~\ref{e:TodaPElimits} we show that the limiting values of the ratio are
\vspace*{-0.4ex}
\[
\lim_{\kappa\to0}\frac{\overline{\PE}}{\E_\tot}=\frac{1}{2},\qquad
\lim_{\kappa\to\infty}\frac{\overline{\PE}}{\E_\tot}=0.
\label{e:peetot}
\]
In particular, at $\kappa=1.3603$, one has $\overline{\PE}/\E_\tot = 4/9$.
For this value of $\kappa$, a SW in the Toda system distributes its energy in the same ratio as a SW in the Hertz system.
This suggests that one could use this value of $\kappa$ to compare SWs in the two systems.
A SW in the Toda lattice with $\kappa = 1.3603$ has a total energy of $\E_\tot=4.8412$.
We thus compare it to the SW in the Hertz system that has the same value of total energy.
The KE and PE profiles for the SW in the two systems are shown in Fig.~\ref{f:peke}.
Beyond this point, the profiles of the SWs in the Toda system vary, while those of the SWs in the Hertz system remain the same, except for a scaling of the SW energy.
Therefore, capturing the profiles of the SWs in the Hertz system at a specific energy is equivalent to capturing its profiles at an arbitrary energy.

\section{Response to a velocity impulse}
\label{s:delta}

After having discussed and compared the properties of SWs in the Toda lattice and the Hertz system in the previous section,
we now discuss how these SWs can be excited from suitable initial conditions.
We first consider the response of an infinite chain to an initial velocity impulse.
That is, we consider the following initial condition (ICs):
\[
y_n = 0\,,\quad \dot y_n = v\,\delta_{k,n},
\qquad \forall n\in\Integer\,,
\label{e:ic}
\]
where $\delta_{k,n}$ is the Kronecker delta.
Without loss of generality, we take $k=0$ (i.e., we apply the impulse at the origin).
The corresponding kinetic energy is $E_\imp= v^2/2$.
{Part of the reason for considering such a scenario is the
  experimental ability to produce such conditions via
  a boundary excitation in the granular
  crystals~\cite{Nesterenko,sen:2008,IOP}.
  While the controlled distributed initialization of such a system is
  not yet within reach, the ability to drive one of the
  boundaries/walls
  of the granular chain to induce an initial velocity has been
 since early on one of the most canonical ways for exciting the SWs.}

\begin{figure}[b!]
\centering
\includegraphics[width=0.485\textwidth]{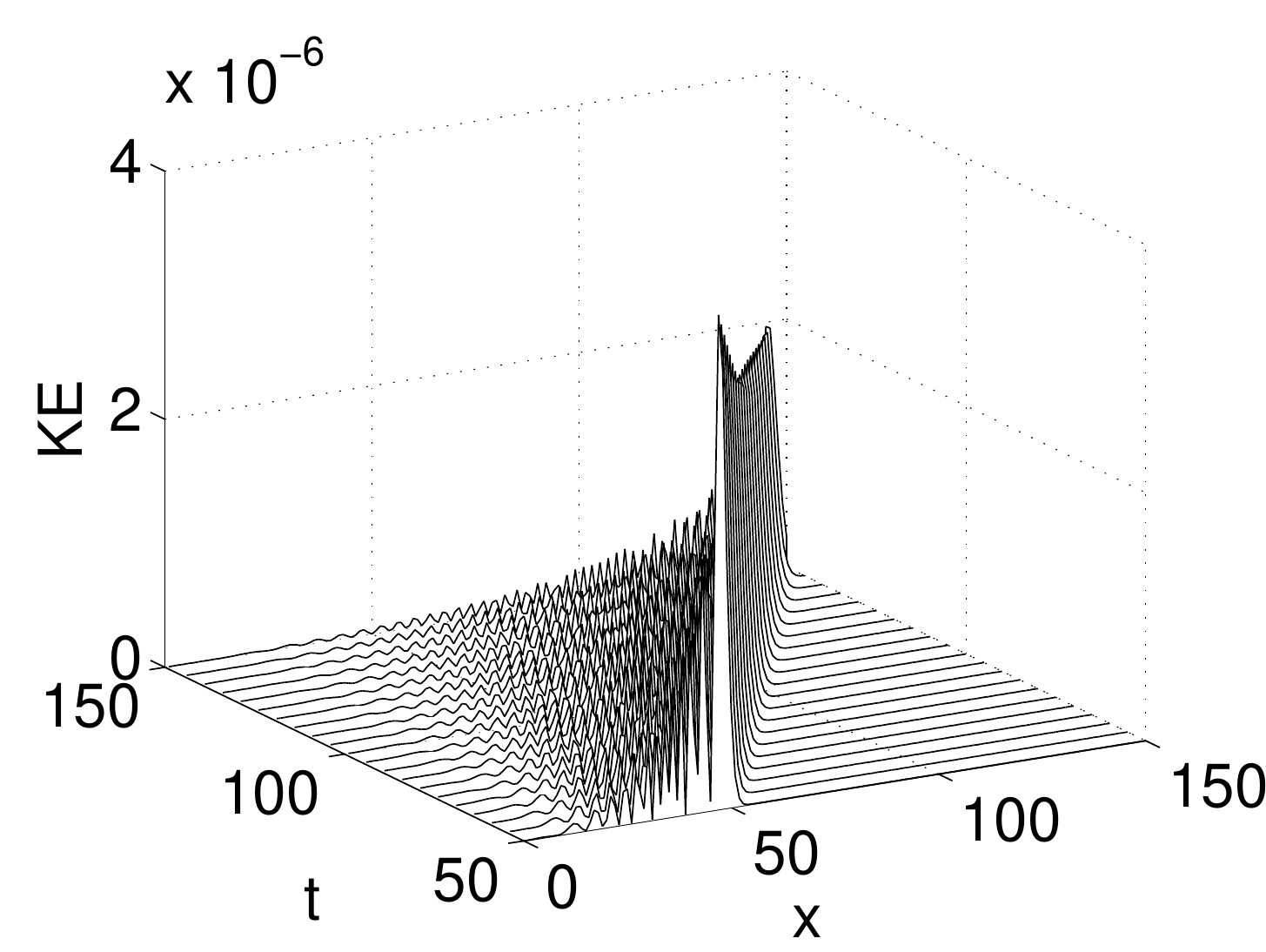}
\includegraphics[width=0.485\textwidth]{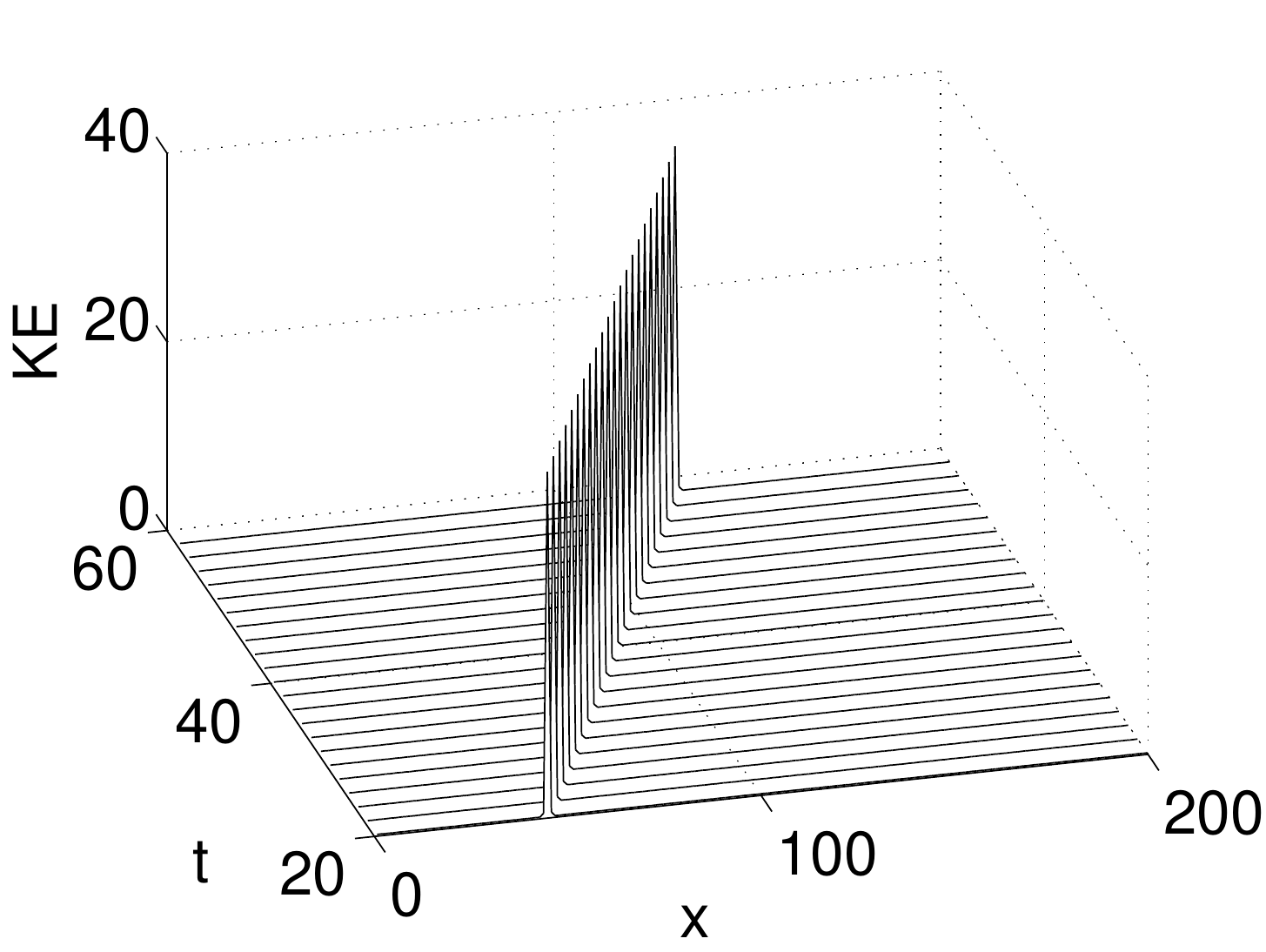}
\kern-\smallskipamount
\caption{The KE of Toda lattice with perturbation $v=0.01$ (left) and $v=10$ (right) respectively.
(To make the figure in the right panel cleaner, the solution was sampled at integer multiples of the temporal period of the oscillations of the KE.)}
\label{f:t}
\end{figure}

Since the Toda potential is superquadratic, it is reasonable to expect that, under suitable conditions, the initial velocity impulse generates one or more SWs.
On the other hand, for small deviations, the Toda potential is approximately quadratic.
Since no SWs are possible in this limit, one could expect that the SWs generated by~\eqref{e:ic} disappear in the limit of weak impulses.
Conversely, as the velocity impulse increases, one can expect the Toda system deviates from linear limit, and
the SW to become more pronounced.
These predictions are borne out by numerical simulations.
Figure~\ref{f:t} shows the response of the Toda system~\eqref{e:toda} under different velocity impules.
The left panel shows the dynamics produced by $v=0.01$,
while the right panel corresponds to $v=10$.
From these figures, it is clear that, with a weak impulse, most of the energy is dispersed,
while with a stronger impulse  the majority of the energy stays localized.
Below we show how one can make the above claims more precise.
To do so, we will use the IST to compute the portion of the impulse energy that goes into the SW, i.e., the quantity $\E_\sw/\E_\imp$,
as a function of an arbitrary impulse velocity $v$.
{We will then compare these findings to the non-integrable
  Hertzian case.}

\subsection{Toda lattice: Integrability and scattering problem}

Using the IST for the Toda lattice, we can compute the soliton parameter $\kappa$ of the SW generated by an arbitrary velocity impulse $v$.
This will allow us to obtain a quantitative characterization of the relation between $\E_\sw/\E_\imp$ and~$v$.

Recall that one can formulate the IST for the Toda lattice using the
Flaschka
variables~\cite{Toda}
\vspace*{-0.4ex}
\[
a_n= \frac{1}{2}\e^{-(y_{n+1}+y_n)/2},\quad
b_n= \frac{1}{2}\dot y_n.
\]
In terms of $a_n$ and $b_n$, the system~\eref{e:toda} is written as
\begin{subequations}
\begin{gather}
\dot{a}_n=a_n(b_n-b_{n+1}),\\
\dot{b}_n=2(a_{n-1}^2-a_n^2).
\end{gather}
\label{e:EOM_Flaschka}
\end{subequations}
In turn, \eref{e:EOM_Flaschka} can be written in matrix form
as the Lax equation
\[
\frac{\d L}{\d t}=BL-LB,
\label{e:EOM_matrix}
\]
where
\begin{gather}
L = \begin{bmatrix}
    \ddots & \ddots  &        &          &        \\
    \ddots & b_{n-1} & a_{n-1}&          &        \\
           & a_{n-1} & b_n    & a_n      &        \\
           &         & a_n    & b_{n+1}  & \ddots \\
           &         &        & \ddots   & \ddots
    \end{bmatrix},
\quad
B = \begin{bmatrix}
    \ddots & \ddots &          &         &        \\
    \ddots & 0        & -a_{n-1} &         &        \\
           & a_{n-1} & 0        & -a_n     &        \\
           &        & a_n      & 0        & \ddots \\
           &        &         & \ddots    & \ddots
    \end{bmatrix}.
\end{gather}
The Lax equation~\eref{e:EOM_matrix} is the compatibility condition of the Lax pair
\begin{subequations}
\label{e:lax_pair}
\begin{gather}
L\phi_{n} =\lambda\phi_{n},
\label{e:lax pair a}
\\
\frac{\d\phi_{n}}{\d t} = B\phi_{n},
\label{e:lax pair b}
\end{gather}
\end{subequations}
or equivalently, in component form,
\begin{subequations}
\label{e:lax_pair2}
\begin{gather}
a_{n-1}\phi_{n-1}+b_n\phi_{n}+a_n\phi_{n+1} =\lambda\phi_{n},
\label{e:lax pair a2}
\\
\frac{\d\phi_{n}}{\d t} =a_{n-1}\phi_{n-1}-a_n\phi_{n+1}.
\label{e:lax pair b2}
\end{gather}
\end{subequations}
Equation~\eref{e:lax pair a} is known as the scattering problem, and~\eref{e:lax pair b} is known as the time evolution equation.
The eigenvalue $\lambda$ is called the scattering parameter, and $\phi_{n}= \phi(n,t,\lambda)$ is the corresponding eigenfunction.

\subsection{Infinite Toda lattice response to a velocity impulse}
\label{ss:infinite}

As usual in the inverse scattering transform, the solitons of the Toda lattice are associated with the discrete eigenvalues of the scattering problem
\cite{Flashchka1,Flashchka2}, while the continuous spectrum is associated with the radiative (dispersive) portion of the solution.
To characterize the fraction of the energy of the initial impulse that goes into the SW, we need to compute the discrete eigenvalues $\lambda$ of~\eref{e:lax pair a}, with $a_n$ and $b_n$ determined by the IC~\eref{e:ic},
which in terms of the Flaschka variables become
\[
a_n=\frac{1}{2}\,\quad b_n=\frac{1}{2}v\,\delta_{0,n},\qquad
\forall n\in\Integer\,.
\label{e:ic1}
\]

It is convenient to express the scattering problem in terms of a modified eigenvalue $z$ given by
\vspace*{-1ex}
\[
\lambda=\frac{z+1/z}{2}.
\label{e:z}
\]
In terms of $z$, the continuous spectrum of~\eref{e:lax pair a} is the unit circle $|z|=1$.
Inserting~\eref{e:ic1} and~\eref{e:z} into~\eref{e:lax pair a},
one can write two sets of linearly independent solutions of the scattering problem as
\begin{subequations}
\begin{gather}
\phi_n = z^n,\quad
\=\phi_n = z^{-n},\qquad n\ge 0\,,
\label{e:n>0}
\\
\noalign{\noindent and}
\psi_n = z^{-n},\quad
\=\psi_n = z^n,\qquad n\le 0\,.
\label{e:n<0}
\end{gather}
\end{subequations}
The symmetries of the scattering problem then imply
\begin{equation}
\=\phi(n,t,z) = \phi(n,t,1/z)\,,\qquad
\=\psi(n,t,z) = \psi(n,t,1/z)\,.
\label{e:symmetries}
\end{equation}
Importantly, $\phi_n$ and $\psi_n$ are analytic in $|z|<1$; while $\=\phi_n$ and $\=\psi_n$ are analytic in $|z|>1$.
These two sets of fundamental solutions are related to each other by the scattering relation,
\vspace*{-1ex}
\begin{equation}
\begin{pmatrix}
    \psi_n \\ \=\psi_n
   \end{pmatrix} =
  S \begin{pmatrix}
    \=\phi_n\\
    \phi_n
   \end{pmatrix},
\qquad n\in\Integer\,,\quad |z|=1\,,
\label{e:scattering}
\end{equation}
where the $n$-independent matrix $S(t,z)$ is is known as the scattering matrix.
Using the symmetries of the problem, one can express $S$ as
\begin{equation}
S(t,z) =
   \begin{pmatrix}
  \alpha(z)  &\   \beta(z) \\
  \beta(1/z)  &\  \alpha(1/z)
  \end{pmatrix}.
\end{equation}
The discrete eigenvalues of the scattering problem correspond to the bound states, and are associated to the zeros of $\alpha(z)$ in $|z|<1$, at which $\psi_n$ and $\phi_n$ are proportional to each other.

\begin{figure}[t!]
\centering
\includegraphics[scale=0.435,trim={0.05cm 0cm 0.8cm 0cm},clip]{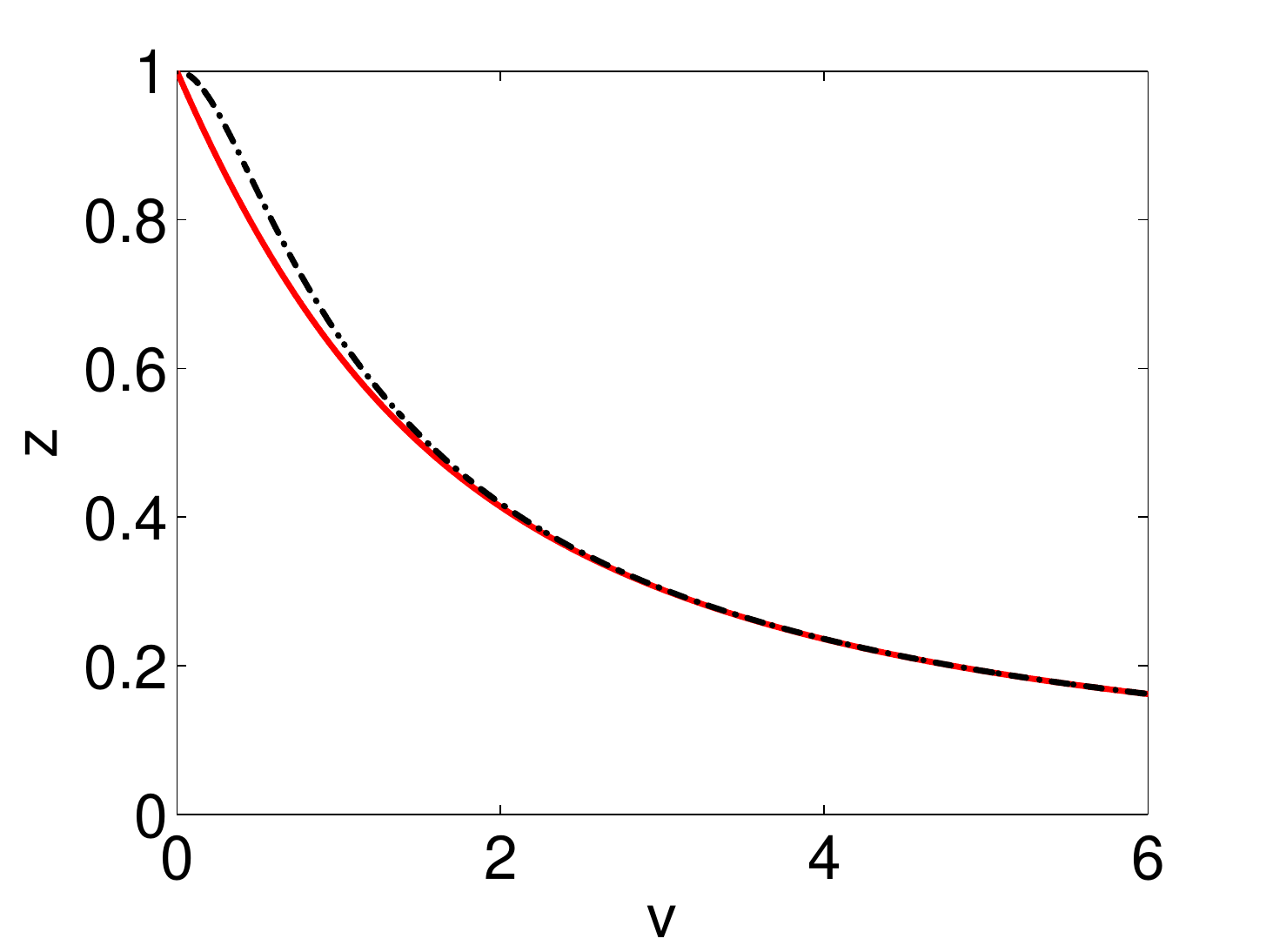}\quad
\includegraphics[scale=0.435,trim={0.15cm 0cm 0cm 0cm},clip]{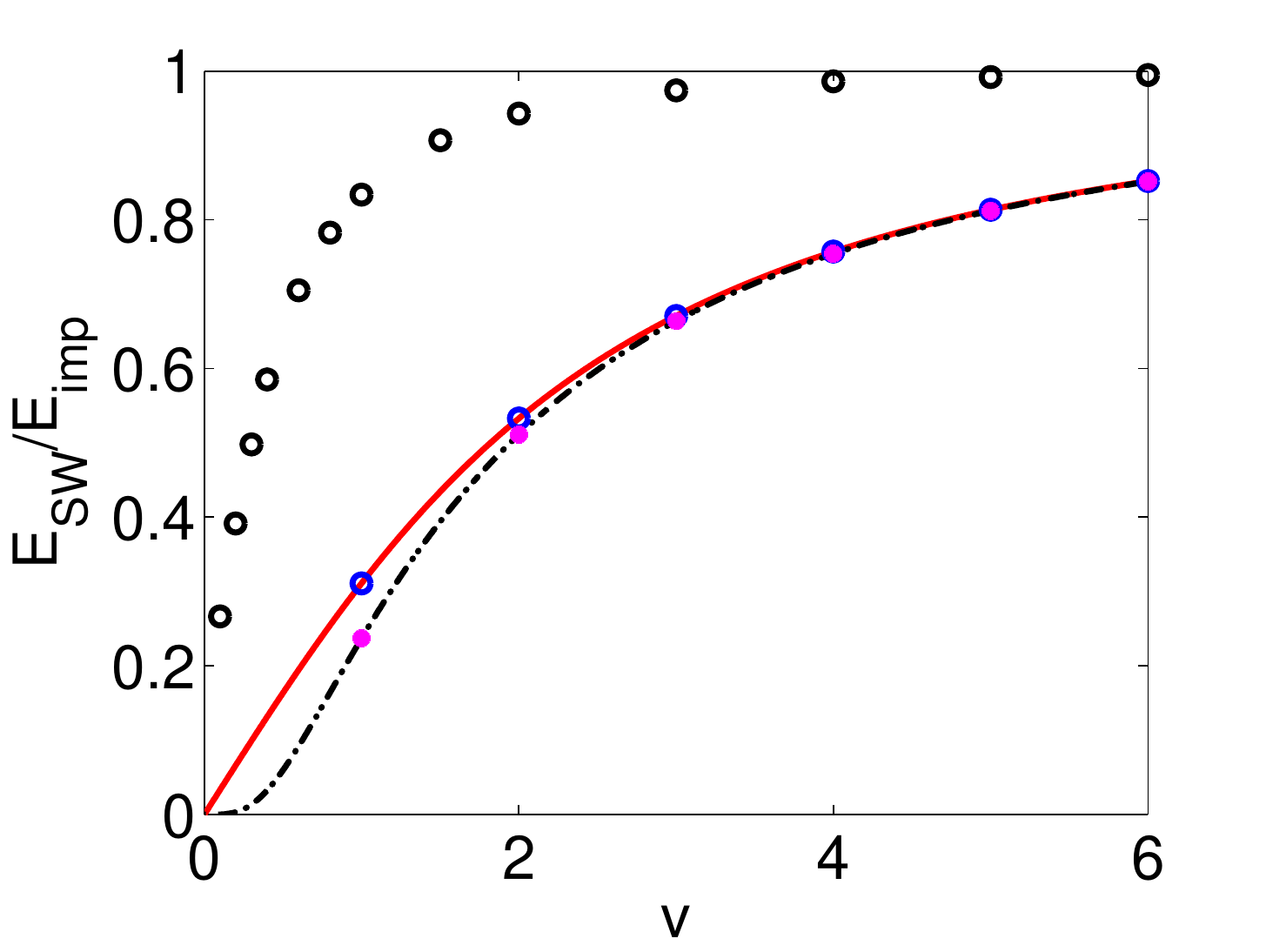}
\kern-\smallskipamount
\caption{Left panel: Discrete eigenvalue $z$ of the scattering problem~\eref{e:lax pair a} as a function of the impulse velocity $v$ for the Toda lattice.
Right panel: the ratio $\E_\sw/\E_\imp$ as a function of the impulse velocity $v$.
In both cases, the solid red lines and dashed black lines represent respectively the analytical results for an infinite chain and a semi-infinite chain case with Dirichlet BC.
The blue circles and magenta stars in the right panel represent respectively numerical results for the infinite chain and for a semi-infinite chain with Dirichlet BC; the black circles represent numerical results for a semi-infinite chain with open BCs (for which no analytical results are available).
}
\label{f:e_sw}
\end{figure}

Evaluating the scattering relation~\eref{e:scattering} at $n=0$ and at $n=1$, we can express the scattering matrix as
\[
S=\begin{pmatrix}
 \psi_0  &\  \psi_1 \\
 \= \psi_0  &\  \=\psi_1
  \end{pmatrix}
  \begin{pmatrix}

  \= \phi_0  &\  \= \phi_1 \\
  \phi_0  &\  \phi_1
  \end{pmatrix}^{-1}.
\]
Using~\eref{e:lax pair a2} and~\eqref{e:n<0} we have
\[
\psi_1 = 1/z-v,\quad \=\psi_1 = z-v.
\label{e:psi}
\]
Combining~\eref{e:n>0},~\eref{e:n<0}, \eref{e:psi} and~\eqref{e:symmetries}, we then obtain
\[
S = \frac1{1-z^2} \begin{pmatrix}
  1-z(v+z)  &\   vz\\
  -vz  &1+z(v-z)
  \end{pmatrix}.
\]
Looking for $\alpha(z)=0$ and solving the resulting quadratic equation, we then obtain
\[
z_{\pm}=\frac12 \big(-v\pm\sqrt{v^2+4} \,\big).
\]
Since we need $|z|<1$, the root $z_-$ is inadmissible.  We therefore have a single discrete eigenvalue $z=z_+$.
The dependence of $z$ on $v$ is shown in the left panel of Fig.~\ref{f:e_sw}. The soliton parameter $\kappa$ relates to $z$ by~\cite{Toda}
\[
\kappa=-\log |z|.
\label{e:kappaz}
\]
Recall~\eref{e:solitonenergy}, we obtain
\[
\frac{\E_\sw}{\E_\imp}=\frac{2(\sinh\kappa\cosh\kappa-\kappa)}{\frac{1}{2}v^2},
\label{e:e_sw}
\]
with $\kappa$ given by~\eref{e:kappaz}.
A comparison between~\eref{e:e_sw} and numerical results is shown in the right panel of Fig.~\ref{f:e_sw}.
We can show that
\[
\lim_{v\to0}\frac{\E_\sw}{\E_\imp}=\frac{1}{3}v+O(v^3),\quad \lim_{v\to\infty}\frac{\E_\sw}{\E_\imp}=1-O(v^{-2}\log v),
\]
i.e., for small perturbations, only negligible amount of the initial
energy goes to the soliton, while for large perturbations, almost all
the initial energy goes to the soliton;
{this is in line with the above reported numerical observations.
  Importantly, however, it is relevant to note that a (single) solitary wave is
  always excited even if bearing an very small fraction of the original
  kinetic energy; i.e., there is no threshold for the excitation of the
  solitary wave in this setting.}

\subsection{Semi-infinite Toda lattice}
\label{ss:semi-infinite}
Next we consider a semi-infinite Toda lattice, and we apply a perturbation to the particle at the end of the chain with a velocity impulse.
Since we are now dealing with an initial-boundary value problem, we need to specify appropriate boundary conditions (BCs) for the problem.
There are two cases of particular physical interest: the one in which the first particle is confined by a wall and that of an open boundary.

We first discuss the case where a wall is placed next to the first particle.
This is equivalent to having a particle with infinite mass before the first particle.
This imaginary particle is subjected to Dirichlet BC, therefore the semi-infinite chain can be oddly extended to an infinite chain with IC given by
\[
y_n=0\,,\quad \dot y_n=v\delta_{1,n}-v\delta_{-1,n}\,,
\qquad \forall n\in\Integer\,.
\label{e:ic2}
\]
Or, in terms of Flaschka variables,
\[
a_n=\frac{1}{2}\,,\quad b_n=\frac{1}{2}v\delta_{1,n}-\frac{1}{2}v\delta_{-1,n}\,,
\qquad \forall n\in\Integer\,.
\label{e:ic3}
\]
The ability to perform these extensions to a doubly infinite Toda chain amounts to the fact that the pure Dirichlet BCs is integrable.
We can now perform a similar analysis as in~\ref{ss:infinite} to compute the discrete eigenvalue of~\eref{e:lax pair a}, obtaining
\[
z_+^{\pm} = \pm\sqrt{\frac{-(1+v^2)+\sqrt{(1+v^2)^2+4v^2}}{2v^2}},\quad z_-^{\pm} = \pm\sqrt{\frac{-(1+v^2)-\sqrt{(1+v^2)^2+4v^2}}{2v^2}}.
\]
Since we need $|z|<1$, only $z_+^{\pm}$ are admissible. By~\eref{e:kappaz}, we see that the soliton parameters correspond to $z_+^{\pm}$ are the same.
Further analysis suggests that $z_+^{\pm}$ correspond to solitons moving with opposite velocity.
This should not be surprising, since~\eqref{e:ic2} implies that we initialize the chain with two equal and opposite velocity impulses.
The value of $z=z_+^+$ as a function of $v$ is shown in the left panel of Fig.~\ref{f:e_sw}.
As before, from $z$ one obtains $\kappa$ and therefore $\E_\sw/\E_\imp$.

No analytical results are available in the case of open BCs, because the first particle obeys a different equation of motion, which breaks
the integrability of the system.
However, a comparison between the analytical values of $\E_\sw/\E_\imp$ and results of numerical simulations is shown in the right panel of Fig.~\ref{f:e_sw}.
As illustrated in the right panel of Fig.~\ref{f:e_sw}, as $v$ increases, $\E_\sw/\E_\imp$ in the open BC case approaches the limiting value 1 much faster than in the other two cases.

\subsection{Comparison with the Hertz system}

In the previous section we have seen that how, using the IST, for the Toda lattice
one can obtain the resulting soliton parameter $\kappa$ (and therefore the soliton velocity) produced by an initial impulse.
No such tool is available for the Hertzian system, however.
We now therefore investigate whether the response of the Hertz system
to an initial velocity impulse results in similar dynamics as in the
Toda lattice.
{Notice that this problem is one closely related to case examples
  that
  have been studied computationally~\cite{sokolow} and even
  experimentally~\cite{melo} previously where a bead at the edge
  of the chain is impacted by a striker of different masses; see also~\cite{Hinch}.}


The Hertz system without precompression (i.e., with $\Delta=0$) is fundamentally different from the Toda lattice in this respect,
since it is intrinsically nonlinear, i.e., it does not admit a linear
limit, {as we discussed above. In that light, the energy cannot be
  distributed to both linear and nonlinear waves, but rather is
  immediately
  ``quantized'' in nonlinear wavepackets, i.e., in SWs}. 
Indeed, as shown in numerical simulations, the leading SW in the Hertz
chain takes more than $99\%$ of the impulse energy; see
also~\cite{Panayotis}. In fact,
our computations indicate that for the Hertz-type system (i.e.,
$\alpha$ in~\eref{e:hertzian} is greater than 2) without
precompression, the leading SW acquires a constant portion of the
impulse energy under arbitrary impulse. The value of $\E_\sw/\E_\imp$
with different values of $\alpha$ is given in
Table~\ref{t:ratio}. {The
  dependence on the interaction exponent $\alpha$ of this
  universal
  (among impulses) feature has not been reported previously to the
  best of our knowledge.}

\begin{table}[t!]
\caption{$\E_\sw/\E_\imp$ with different values of $\alpha$ for the Hertz system~\eqref{e:hertz}.}
\medskip
\centering
\begin{tabular}{c|ccccccc}
$\alpha$         & 2.1 & 2.2 & 2.3 & 2.4 & 2.5 & 3 & 4\\
\hline
$\E_\sw/\E_\imp$ $\%$ & 97.60 & 98.35 & 98.82 & 99.14 & 99.37 & 99.85 & 99.99\\
\end{tabular}
\label{t:ratio}
\end{table}

{On the other hand, as discussed above, it is possible to consider
  the Hertz and the Toda systems on more proximal footing, when
  considering
  in the former the case with precompression.}
The Hertz system with precompression does admit a linear limit, like the Toda system.
Therefore, in this case we expect that $\E_\sw/\E_\imp$ increases as $v$ increases. However, no satisfactory analytical tools are available to characterize such a system beyond the linear limit. So in this case we will obtain $\E_\sw/\E_\imp$ as a function of $v$ numerically.

For the purpose of comparing the Hertz system to the Toda system, it
is useful to rescale $y$ and $t$ in~\eref{e:hertz} so that the
equations of motion in these two systems agree up to the square order
as was suggested in the work of~\cite{Shen}. To do so, we introduce dimensionless variables $\xi_n$ and $\tau$ such that $y_n=\xi_ny_0$ and $t=\tau t_0$ with $y_0$ and $t_0$ arbitrary for now. Then~\eref{e:hertz} becomes (taking $\alpha = 2.5$ for definitiness)
\vspace*{-1ex}
\begin{equation}
\label{e:hertz_non}
\begin{aligned}
\frac{\d^2\xi_n}{\d \tau^2} = &\frac{4}{15} c\Delta^{1/2}t_0^2[(\xi_{n+1}-\xi_n)-(\xi_n-\xi_{n-1})]\\
                          &-\frac{15}{16}c\Delta^{-1/2}y_0t_0^2[(\xi_{n+1}-\xi_n)-(\xi_n-\xi_{n-1})^2]+\dots
\end{aligned}
\end{equation}
Setting $t_0=\sqrt{4/({15c\Delta^{1/2}})}$ and $y_0=2\Delta$, we then obtain a system that agrees with the Toda system~\eref{e:toda} up to second order in the displacement difference.
The IC corresponding to~\eref{e:ic} is given by
\vspace*{-1ex}
\[
\xi_n=0\,, \quad \frac{\d \xi_n}{\d\tau}=\frac{t_0}{y_0}v\delta_{k,n}\,, \qquad\forall n\in\Integer\,.
\label{e:ic_hertz}
\]

Note that the coefficients of~\eref{e:hertz_non} do not depend on $\Delta$, but the rescaled IC~\eref{e:ic_hertz} depends on $\Delta$. Specifically the rescaled velocity $\d \xi_n/\d\tau$ scales like $\Delta^{-5/4}$. Therefore the results are independent of precompression as long as one rescales the IC accordingly.

For the Hertz system with precompression, one does not have $\phi(0)=0$,
unlike what happens in the Toda lattice.
So, if one used the potential $\phi(r)$ to compute energies, the energy corresponding to the velocity impulse~\eqref{e:ic} would not simply be $v^2/2$.
Therefore, for the precompressed Hertz system, one must replace $\phi(r)$ with
\begin{equation}
\begin{aligned}
\tilde\phi(r)=&\phi(r)+\alpha c\Delta^{\alpha -1}\,r-c\Delta^\alpha,
\end{aligned}
\end{equation}
where $c$ and $\alpha$ are the constants appearing in~\eqref{e:hertz}.
for an infinite lattice, the potential $\tilde\phi(r)$ yields the same equation of motion~\eref{e:EOM} as $\phi(r)$
However, unlike $\phi(r)$, $\tilde\phi(r)$ satisfies the constraints $\tilde\phi'(0)=\tilde\phi(0)=0$
as required in \cite{FrieseckeWattis}.
The energy of a SW in a precompressed Hertz system is then \cite{MacKay}
\vspace*{-0.6ex}
\[
E_\sw=\sum_n\bigg(\frac{1}{2}m\dot{y}_n^2+\frac{1}{2}(\tilde\phi(y_n-y_{n-1})+\tilde\phi(y_{n+1}-y_n))\bigg),
\]
where the summation runs over all the particles involved in the SW.

\begin{figure}[t!]
\centering
\includegraphics[scale=0.435,trim={0.05cm 0cm 0.8cm 0cm},clip]{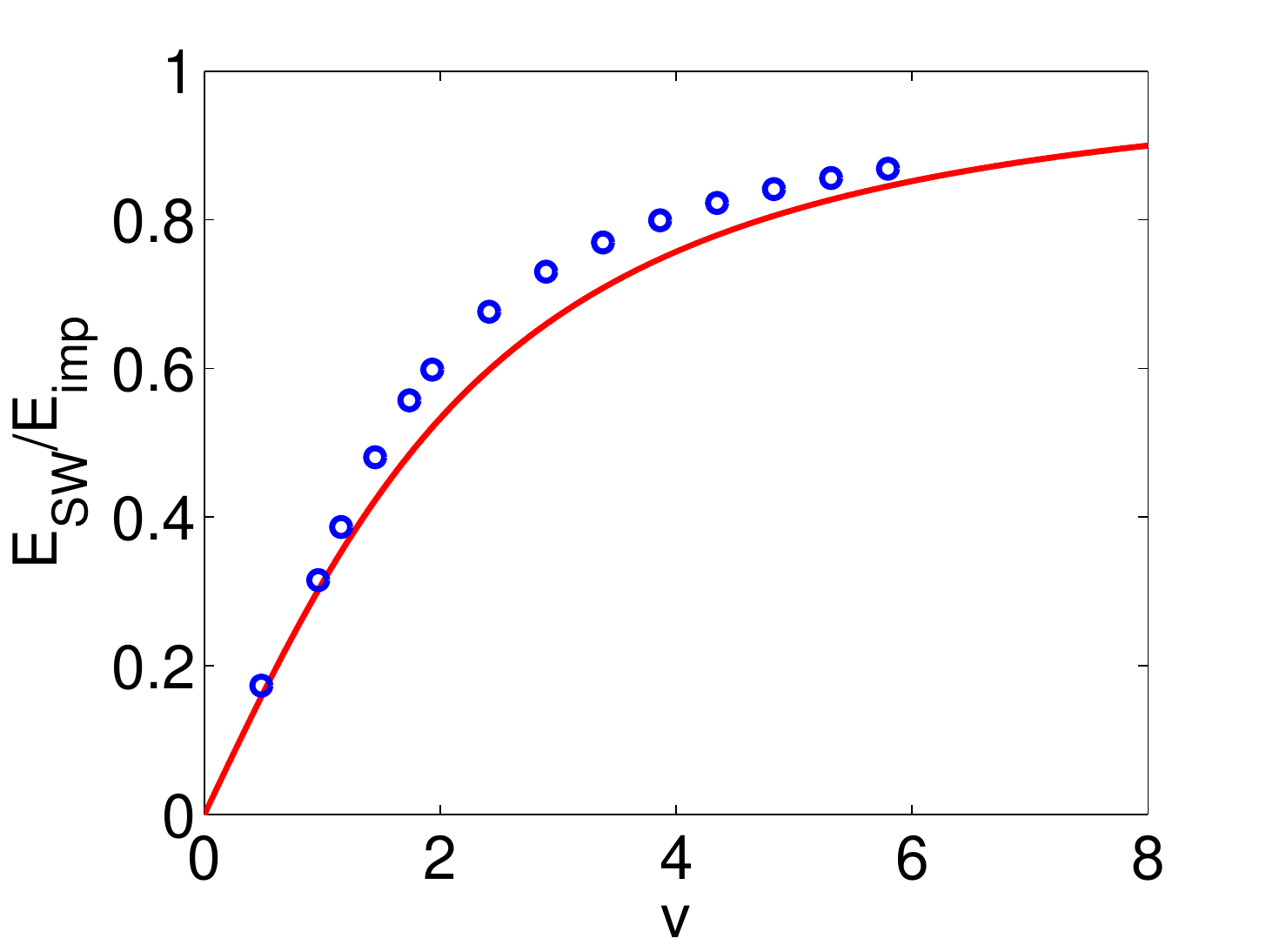}
\includegraphics[scale=0.435,trim={0.15cm 0cm 0cm 0cm},clip]{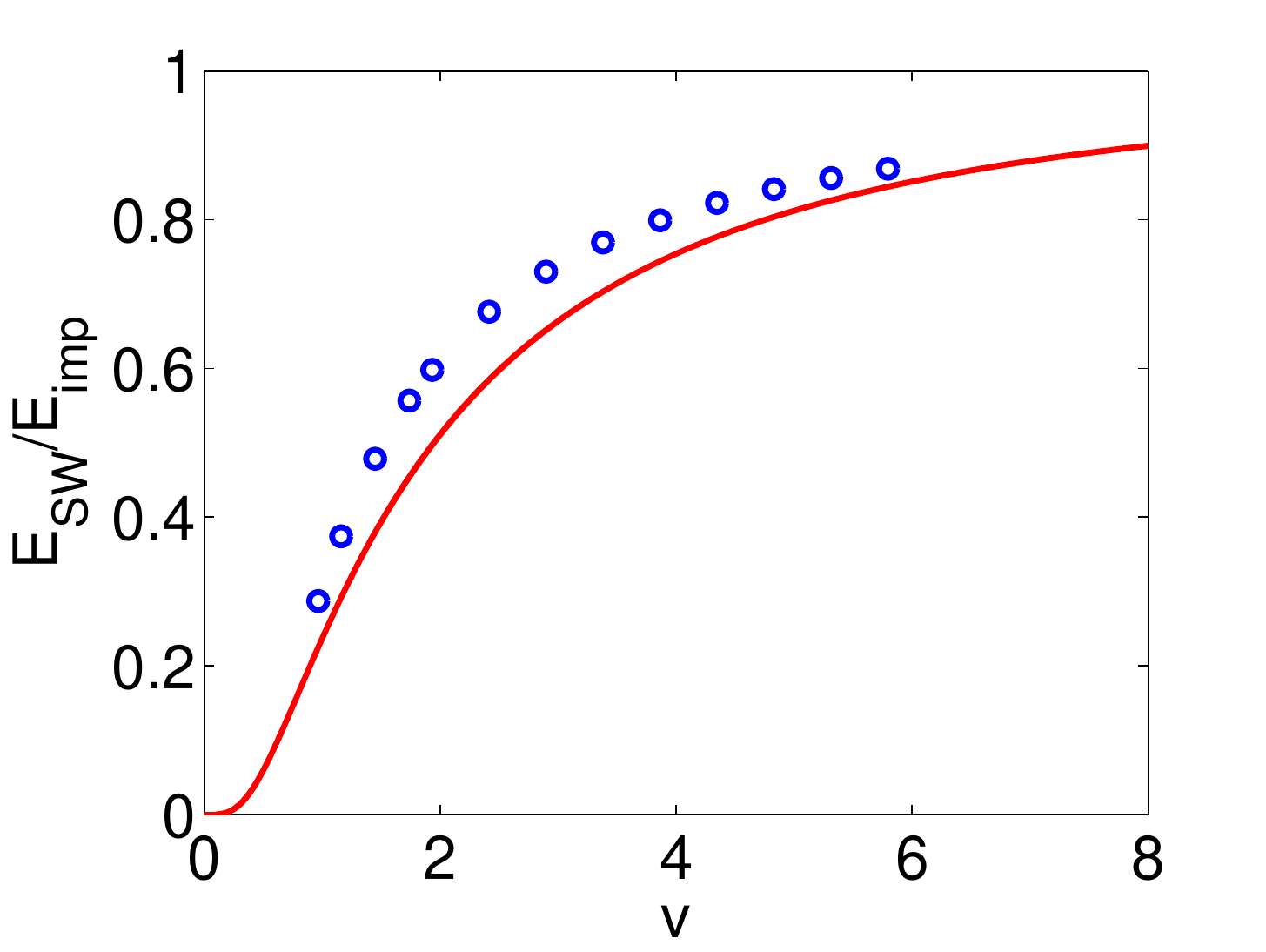}
\kern-\smallskipamount
\caption{A comparison of $\E_\sw/\E_\imp$ vs $v$ between the precompressed Hertz (blue dots) and the Toda systems (red line).
For the Hertz system, $v$ is the rescaled velocity $\d \xi_n/\d\tau$ in~\eref{e:ic_hertz}.
The left panel represents the case of an infinite chain, while the right panel represents the case of a semi-infinite chain with fixed BC.}
\label{f:ht1}
\end{figure}

Figure~\ref{f:ht1} shows the dependence of the ratio $\E_\sw/\E_\imp$ on the impulse velocity $v$ in both the precompressed Hertz system and the Toda system.
As can be seen from the figure, the behavior of both systems is qualitatively the same, both in the case of an infinite chain and in the case of a
semi-infinite chain.
Therefore we can conclude that the precompressed Hertz system, which admits a linear limit, demonstrates similar dynamics in response to a velocity impulse
as the Toda system.

We should point out that numerical measurements with small impulse velocity $v$ are difficult in both the Toda and Hertz systems, because small SWs take a long time to separate from the radiation. Nonetheless, in such a regime, the two systems behave similarly, as the displacement of particles is relatively small and hence the contributions of higher orders in~\eref{e:hertz_non} are negligible. 
For the Toda system, we have an analytical prediction, which is valid for all values of impulse velocity $v$. 
Hence one can use the analytical predictions in the Toda lattice to characterize the behavior of the Hertz system in this regime.

\section{Summary and discussion}
\label{s:discussion}

In conclusion, we have shown that the behavior of the KE and the PE carried by a SW in both Toda and Hertz systems is oscillatory in time, and that
this oscillatory behavior is due to the {lattice shift invariance
  as the solitary wave travels from site to site.
  Importantly, we obtained analytical closed form expressions
  for these dependencies, showcasing that these oscillations can be
captured by cnoidal functions in the Toda system; a corresponding
sinusoidal functional form was numerically illustrated
in the Hertz system.}
We also showed that the shape of the KE and PE of these systems as a function of $n$ are similar to each other at $\E_\tot=4.84$,
and that beyond this point the profiles in the Hertz system remain the
same except for the overall scaling of the SW energy. {More
  generally,
  we compared the solitary waves of the Toda and the Hertz system.
  In the presence of precompression and in the vicinity of the linear
  limit, the two can be seen to be rather proximal. In the highly
  nonlinear
  cases, while similarities exist (e.g. in comparable dependencies
  of the amplitude vs. energy of the solitary wave), there also
  exist significant differences, including the width of the waves
  and the nature of the decay tails in the different cases.}

We then studied the response of the infinite Toda lattice to an initial velocity impulse.
By employing the IST we showed how one can obtain the ratio between the energy of the resulting SW and the energy of the impulse,
and how this ratio depends on the impulse velocity.
{We corroborated these exact analytical results with numerical simulations}.
We then studied the response of a semi-infinite Toda lattice in which the first particle is perturbed with a velocity impulse.
In this case, the results depend on the specific boundary conditions (BCs) considered.
For fixed BCs we introduced an odd extension of the semi-infinite chain and we obtained an analytical expression of the above ratio by employing the IST.
For ``free'' BCs, however, no such approach was available, and we
resorted to numerical simulations. 

For the Hertzian chain with zero precompression, {following up on
  earlier
  work, we confirmed that 
  more than 99\% of the energy in an arbitrary initial impulse goes
  to the resulting SW and discussed the independence of this finding
  from the size of the impulse. Moreover, we explored how this
  fraction
  is modified for different nonlinear force exponents $\alpha$
  indicating
  a weak monotonically increasing dependence (once again independent
  of the impulse size $v$) on $\alpha$.}
{We then extended the relevant comparison to the 
Hertz chain with precompression; there, the} percentage of energy that goes to solitary wave exhibits a similar behavior as in the Toda system.
We then performed a quantitative comparison of the values of this ratio in the Toda and precompressed Hertz systems by nondimensionalizing the precompressed Hertz system, and rescaling its time and position coordinates so that the equation of motion in the precompressed Hertz system agrees with that in the Toda system up to second order in the displacement difference.
Our comparison shows that, at relatively small impulse energies, the ratios in the two systems are close to each other;
however, as the impulse energy is increased, the ratios in these systems deviate, even though they both eventually approach the same limiting unit value.

The results of this study open a number of interesting theoretical questions.
For example:
\\[0.4ex]
\hglue1em
(a) We demonstrated numerically that the temporal oscillations of the KE and the PE in the Hertz system are described by trigonometric functions at all energies.
However, whether this dependence can be proven mathematically is still
an open question. {It is relevant to note that this is a nontrivial
task as there is no closed functional form for the SW in the highly
nonlinear Hertzian case but only approximate ones~\cite{SenManciu,yuli}.}
\\[0.4ex]
\hglue1em
(b) Similarly, it would be desirable to have a rigorous proof of the fact that the fraction of the energy that goes to the SW in the
Hertz system is independent of the energy of the initial impulse.
\\[0.4ex]
\hglue1em
(c) An analytical characterization of the semi-infinite Toda lattice with open BCs and a velocity perturbation at the first particle is still absent.
Note that open BCs are much more complicated to deal with than fixed
BC mathematically, since they break the integrability of the lattice.
\\[0.4ex]
\hglue1em
(d) Another interesting open topic is a detailed characterization of the radiation generated by a velocity impulse in the Toda lattice,
which could be obtained by approprate use of the IST as was done in other discrete and continuous integrable systems
(e.g., see~\cite{AS1981}).
{More generally, the vein of ``contact'' between the Toda chain
  and the granular chain with precompression can be used for a variety
  of
  semi-analytical efforts. One can explore different types of initial
  conditions
  tailored towards exciting multiple solitary waves and exploring
  their
  interactions in the Toda
  lattice and by extension in the granular precompressed chain.
  Another vein important for future work concerns the connection of
  dispersive shock waves in the Toda lattice (see, e.g.,~\cite{bloch})
  to the study of analogous features in the Hertzian chain (for a
  discussion of the latter see, e.g.,~\cite{choong}).
  Such directions are currently  under consideration and will be reported
in future studies.}

\section*{Acknowledgments}

This work was partially supported by the National Science Foundation under grant numbers DMS-1614623 and DMS-1615524.

\bigskip
\input appendix3a

\bigskip
\input references3a


\end{document}

%% file: appendix3a.tex
\def\Ai{\mathop{\rm Ai}\nolimits}
\def\Bi{\mathop{\rm Bi}\nolimits}
\def\PE{\mathrm{PE}}
\def\KE{\mathrm{KE}}
\setcounter{section}0
\def\thesection{Appendix~\Alph{section}}
\def\theequation{\Alph{section}.\arabic{equation}}
\def\thetheorem{\Alph{section}.\arabic{theorem}}
\def\thefigure{\Alph{section}.\arabic{figure}}

\section{Oscillation of the PE in the Hertzian system at different energies}
\label{a:profile}

As we discussed in section~\ref{s:KEPE},
the PE and KE of SWs in the Hertzian system also demonstrate oscillatory behavior,
like the SWs in the Toda system.
Here, we show that the average value and oscillation amplitude are proportional to the overall energy.

Let us denote by $P(x - v_\sw t;E)$ the spatial profile at time $t$ of the PE of a SW in the Hertzian system with total energy $E$.
(Note that the temporal variation amounts to a continuous variation of the argument, cf.~\eqref{e:sw}.)
Without loss of generality, we can take the peak of $P(x;E)$ to be located at $x=0$.
It should be clear that the temporal evolution of the PE is simply equivalent to a translation of the spatial variable $x$, and that
its maximum and minimum values occur respectively at $t=0$ and $t = 1/(2v_\sw)$.
\begin{subequations}
\begin{gather}
\overline{\PE}(E) = \frac{1}{2}\sum_{n\in\Integer} (P(n;E) + P(n/2;E))\,,\label{e:average}\\
\Delta \PE(E) = \sum_{n\in\Integer} (P(n;E) - P(n/2;E))\,.
\end{gather}
\end{subequations}
Recalling that the profile of SWs in the Hertzian system is universal, and that only its amplitude depends on $E$
[cf.\ \eqref{e:sw}],
we then have
\[
P(x;{a E})= h(a)\,P(x;E)\,,
\label{e:profilepe}
\]
where $h(a)$ is a yet-to-be-determined function
(see also the discussion in section~\ref{s:vsw}).
This relation
immediately implies
\[
\overline{\PE}(aE) = h(a)\,\overline{\PE}(E).
\nonumber
\]
Note, however, that the virial theorem also implies
\bse
\[
\overline{\PE}(aE) = a\,\overline{\PE}(E).
\label{e:averagepe2}
\]
Combining the last two equations, we then immediately have $h(a) \equiv a$.
Thus, $P(x;a E)=a\,P(x;E)$.
which also implies
\[
\Delta\PE(aE) = a\, \Delta\PE(E)\,.
\]
\ese

\section{Limiting behavior of $\overline{\PE}/{\E_\tot}$ in the Toda system as $\kappa\to0$ and $\kappa\to\infty$}
\label{e:TodaPElimits}

Recall~\eref{e:ellipticfit} and~\eref{e:peavg}, we have
\begin{equation}
\begin{aligned}
\overline{\PE}&=
                -2\kappa+\sinh^2\kappa\bigg(\frac{2K'_m}{\pi}\bigg)^2\bigg(\frac{E'_m}{K'_m}-m\bigg)+\frac{\sinh^2\kappa}{2K_m}\bigg(\frac{2K_m'}{\pi}\bigg)^2m\int_0^{2K_m}\cn^2(t;m)\d t\\
              &=\PE_{\min}+\frac{4\sinh^2\kappa}{\pi^2} K_m'^2\bigg(\frac{E_m}{K_m}-1+m\bigg)
\equiv \PE_{\min}+\PE_{\integral}
\end{aligned}
\end{equation}
and also recall that $\kappa$ is related to $m$ by~\eref{e:kappam}.

We first discuss the limit $\kappa\to0$. Note that in this limit
\begin{subequations}
\begin{gather}
\kappa=-\frac{\pi^2}{\log\frac{m}{16}}+O\bigg(\frac{m}{\log^2 m}\bigg),\\
\E_{\tot}=\frac{4}{3}\kappa^3+\frac{4}{15}\kappa^5+O(\kappa^7),
\end{gather}
\end{subequations}
as a result we have
\[
\E_{\tot}=-\frac{4}{3}\bigg(\frac{\pi^2}{\log\frac{m}{16}}\bigg)^3\bigg[1+O\bigg(\frac{1}{\log^2 m}\bigg)\bigg].
\label{e:etotm}
\]
Also note that
\begin{subequations}
\begin{gather}
\sinh^2\kappa = \bigg(\frac{\pi^2}{\log\frac{m}{16}}\bigg)^2+\frac{1}{3} \bigg(\frac{\pi^2}{\log\frac{m}{16}}\bigg)^4
+O\bigg(\frac{1}{\log^6 m}\bigg),\\
\bigg(\frac{2K'_m}{\pi}\bigg)^2\bigg(\frac{E_m}{K_m}-1+m\bigg)=O(m\log^2 m),\\
\bigg(\frac{2K'_m}{\pi}\bigg)^2\bigg(\frac{E'_m}{K'_m}-m\bigg)=-\frac{2}{\pi^2}\log\frac{m}{16}+O(m\log^2 m),
\end{gather}
\end{subequations}
therefore
\[
\overline{\PE}=-\frac{2}{3}\bigg(\frac{\pi^2}{\log \frac{m}{16}}\bigg)^3\bigg[1+O\bigg(\frac{1}{\log^2 m}\bigg)\bigg].
\label{e:peavgm}
\]
Combining~\eref{e:etotm} and~\eref{e:peavgm}, we have
\[
\lim_{\kappa\to0}\frac{\overline{\PE}}{\E_\tot}=\frac{1}{2}+O(\kappa^2).
\]
Next we consider the limit $\kappa\to\infty$, in this limit we have
$m\to1$, we then write $m=1-m'$. Following a similar discussion, we have
\begin{subequations}
\begin{gather}
\kappa=\log\frac{16}{m'}+O(m'),\\
\lim_{\kappa\to\infty}\frac{\overline{\PE}}{\E_\tot}=O\bigg(\frac{1}{\log m'}\bigg).
\end{gather}
\end{subequations}
Therefore
\[
\lim_{\kappa\to\infty}\frac{\overline{\PE}}{\E_\tot}=O(1/\kappa).
\]

\section{Scaling between $v_\sw$ and $v$ for different impulses}
\label{s:vsw}

The discussion in the previous sections raises the obvious question of what is the precise dependence of the velocity $v_\sw$
of the resulting SW on the velocity $v$ of the impulse for the Hertzian chain.
Numerical simulations suggest that, for the Hertz chain, one has $v_\sw\sim v^{0.2}$~\cite{sen:1998},
while for a quartic lattice, one has $v_\sw\sim v^{0.5}$~\cite{Mahan:2008}.
Indeed, one can show that, for lattices with Hertz-type potentials given by~\eref{e:hertzian}
with arbitrary values of $\alpha>2$ and $\Delta=0$, {and for two
  distinct
  initial impulses $v_1$ and $v_2$,} one can show that the following relation holds by using scaling analysis~\cite{sen:2008}:
\vspace*{-0.6ex}
\[
v^{(v_2)}_\sw=(v_2/v_1)^{1-2/\alpha}v^{(v_1)}_\sw.
\label{e:v_sw}
\]
In this section we provide an alternative proof by using the Virial theorem.

Recalling the expression~\eqref{e:sw} for the shape of the SW,
let us denote by $Y^{(v)}(n-v^{(v)}_\sw t)$ the displacement of a SW resulting from an impulse $v$.
Similarly, let us denote the KE and PE of the SW by $\KE^{(v)}$ and $\PE^{(v)}$ respectively,
Explicitly,
\vspace*{-1ex}
\begin{subequations}
\label{e:kepesw}
\begin{align}
\PE^{(v)} &= c \sum_n |Y^{(v)}(n+1-v^{(v)}_\sw t) -Y^{(v)}(n-v^{(v)}_\sw t)|^{\alpha}\,,
\label{e:kepesw1}\\
\KE^{(v)} &= \frac{1}{2}\sum_n\bigg(\frac{\d Y^{(v)}(n-v^{(v)}_\sw t)}{\d t}\bigg)^2,
\end{align}
\end{subequations}
where $c$ is the constant in~\eref{e:hertzian}.

Now recall that in the Hertz-type system, a SW takes a constant portion of the impulse energy under arbitrary impulse~$v$.
The virial theorem \cite{goldstein} immediately implies that
\vspace*{-1ex}
\begin{subequations}
\begin{align}
\overline{\PE}^{(k v)}=k^2\overline{\PE}^{(v)},
\label{e:pe}\\
\overline{\KE}^{(k v)}=k^2\overline{\KE}^{(v)}\,.
\label{e:ke}
\end{align}
\end{subequations}
Also recall that the width of the displacement profiles of SWs in Hertz-type systems is independent of the energy of the SW~\cite{SenManciu}.
We therefore have
\vspace*{-1ex}
\[
Y^{(k v)}=g(k)\,Y^{(v)}(n-v^{(k v)}_\sw t),
\]
where $g(k) = A^{(k v)}/A^{(v)}$ is a function yet to be determined.
Combining~\eref{e:kepesw1} and~\eref{e:pe}, however, we immediately have
\vspace*{-1ex}
\[
g(k) = k^{2/\alpha}\,.
\]
From~\eqref{e:kepesw} and combining the above results, we then have
\begin{equation}
\KE^{(kv)}
           = \frac{1}{2}k^{\frac{4}{\alpha}}\sum_n\bigg(\frac{\d Y^{(v)}(n-v^{(kv)}_\sw t)}{\d t}\bigg)^2
           = \frac{1}{2}k^{\frac{4}{\alpha}}(v^{(kv)}_\sw)^2\sum_n\bigg(\frac{\d Y^{(v)}(u)}{\d u}\bigg)^2\,.
\end{equation}
Evaluating this expression for $k=1$ and comparing the two results we then have
\begin{equation}
\KE^{(kv)}  = k^{\frac{4}{\alpha}}\frac{(v^{(kv)}_\sw)^2}{(v^{(v)}_\sw)^2}\KE^{(v)}.
\end{equation}
Recalling~\eref{e:ke}, we then finally obtain~\eqref{e:v_sw}.